\documentclass[aps,notitlepage,superscriptaddress,floatfix,pra,twocolumn,nofootinbib]{revtex4-1}

\usepackage{lipsum}
\usepackage{graphicx}
\usepackage{amsmath,amssymb}
\usepackage{braket}
\usepackage{mathtools}
\usepackage{hyperref}
\usepackage{multirow}
\usepackage{color}
\usepackage{xcolor}
\usepackage[normalem]{ulem}
\usepackage{amsfonts}
\usepackage{float}
\usepackage{pdfpages} % include appendix
\makeatletter
\usepackage{subfigure}
\usepackage{soul, xcolor}
\usepackage{dsfont}

\setstcolor{red}

\newcommand{\dd}{\mathrm{d}}

\def\beq{\begin{equation}}
\def\eeq{\end{equation}}
\def\bea{\begin{eqnarray}}
\def\eea{\end{eqnarray}}

\definecolor{linkblue}{RGB}{35,75,170}
\definecolor{citeblue}{RGB}{45,90,195}
\definecolor{urlblue}{RGB}{55,80,110}

\hypersetup{
    colorlinks=true,
    linkcolor=linkblue,
    citecolor=citeblue,
    urlcolor=urlblue
}

\usepackage{pdfpages} % include appendix
\makeatletter
\AtBeginDocument{\let\LS@rot\@undefined}
\makeatother

\makeatletter

\def\ltxu@dotsep{1.5em}
\def\l@f@section{%
  \addpenalty{\@secpenalty}%
  \addvspace{0.6em}%
}

\renewcommand{\l@subsubsection}[2]{}

\makeatother

\begin{document}

\title{Bayesian Tracking of a Diffusing Target in Two and Three Dimensions}

\author{Ewan McCulloch}
\affiliation{Laboratoire de Physique de l'École Normale Supérieure, CNRS, ENS \& Université PSL; 24 rue Lhomond, 75005 Paris, France}
\author{Adam Nahum}
\affiliation{Laboratoire de Physique de l'École Normale Supérieure, CNRS, ENS \& Université PSL; 24 rue Lhomond, 75005 Paris, France}

\begin{abstract}
We study Bayesian tracking of a diffusing target monitored by a noisy distributed sensor array. Building on an earlier mapping to KPZ growth with a moving defect (or an equivalent directed polymer pinning problem) we determine the phase structure, beyond the previously-studied one-dimensional case, for both Bayes-optimal and suboptimal inference. In \(d=2\), theoretical analysis and numerical simulations both give a depinning transition between a successful tracking phase and a failure phase. Weak-coupling RG shows that Bayes-optimal tracking is always successful in \(d=2\), but failure can arise from overconfident (suboptimal) inference. In \(d=3\), tracking can succeed, or can fail in two distinct ways: the posterior probability distribution may delocalize (no detection), or may become sharply localized, but at the wrong position (a false detection). The two possibilities correspond to Edwards-Wilkinson or Kardar-Parisi-Zhang statistics for the log-posterior. The three  phases meet at a Nishimori-like multicritical point on a Bayes-optimal line in a two-parameter  phase diagram. (Model misspecification alone can drive depinning into either unpinned phase: underconfidence gives diffuse failure, while overconfidence gives localized-but-wrong failure.) We analyze the  transitions between the various phases numerically and with renormalization group arguments. We show that some of these  have  unusual critical behavior, which the conventional $\epsilon$ expansion fails to describe. Recent rigorous results for directed polymers indicate an alternative scenario. Many of  our results, including a scaling relation for exponents at pinning transitions and results for RG flows, are relevant to other phase transitions that involve surface growth or directed polymers in 2+1D or 3+1D.
\end{abstract}

\maketitle

\makeatletter
\renewcommand{\l@subsubsection}[2]{}
\makeatother

\makeatletter
\begingroup
\let\addcontentsline\@gobblethree
\section{Introduction}
%\label{sec:intro}
\endgroup
\makeatother

Hidden information about the state of a  dynamical system may be recoverable from a noisy measurement sequence using Bayesian filtering 
\cite{welch1995introduction,
rabiner2003introduction,
shah2011rumors,offer2018phase,jin2022kardar,Kim2025,Gerbino2025,NahumBayesianCriticalPoints,Gopalakrishnan2026,2026arXiv260423346J,Gerbino2025}.
Conditioning on the measurement record updates a prior distribution over possible histories to a posterior distribution
that reflects our improved knowledge. 
As the precision of monitoring is decreased, we may transition from a phase in which the hidden information can be recovered to a phase where it cannot.
The large-scale properties of monitored many-body dynamics are described by renormalization group (RG) fixed points whose universal properties may differ from those of any unmonitored system \cite{NahumBayesianCriticalPoints,Gopalakrishnan2026}.
An example is the charge-sharpening transition in the dynamics of a classical or quantum system with a conservation law, where measurements can either succeed or fail in determining the system's total charge \cite{NahumBayesianCriticalPoints,Gopalakrishnan2026,PhysRevX.12.041002,PRXQuantum.5.020304,PhysRevLett.129.120604,2026PhRvB.113e4305S,2025PhRvB.112f4304G,2025PhRvB.111b4204P}. A range of measurement or inference-related phase transitions have been studied in classical~\cite{NahumBayesianCriticalPoints,Gopalakrishnan2026,2026arXiv260423346J} and quantum~\cite{PhysRevB.100.134306,PhysRevX.9.031009,PhysRevB.99.224307,PhysRevX.10.041020,PhysRevX.14.041012,zabalo2022operator,nahum2020entanglement,ippoliti2021entanglement,ippoliti2021postselection,jian2023measurement,fava2023nonlinear,tiutiakina2025field,PhysRevLett.132.110403,2022PhRvL.128a0605M,2021PhRvL.126q0602A,loio2023purification,garratt2023measurements,bao2024finite,ferte2024solvable,ferte2026decoherent} many-body systems.

\begin{figure}[t]
    \centering
    \includegraphics[width=1.0\linewidth]{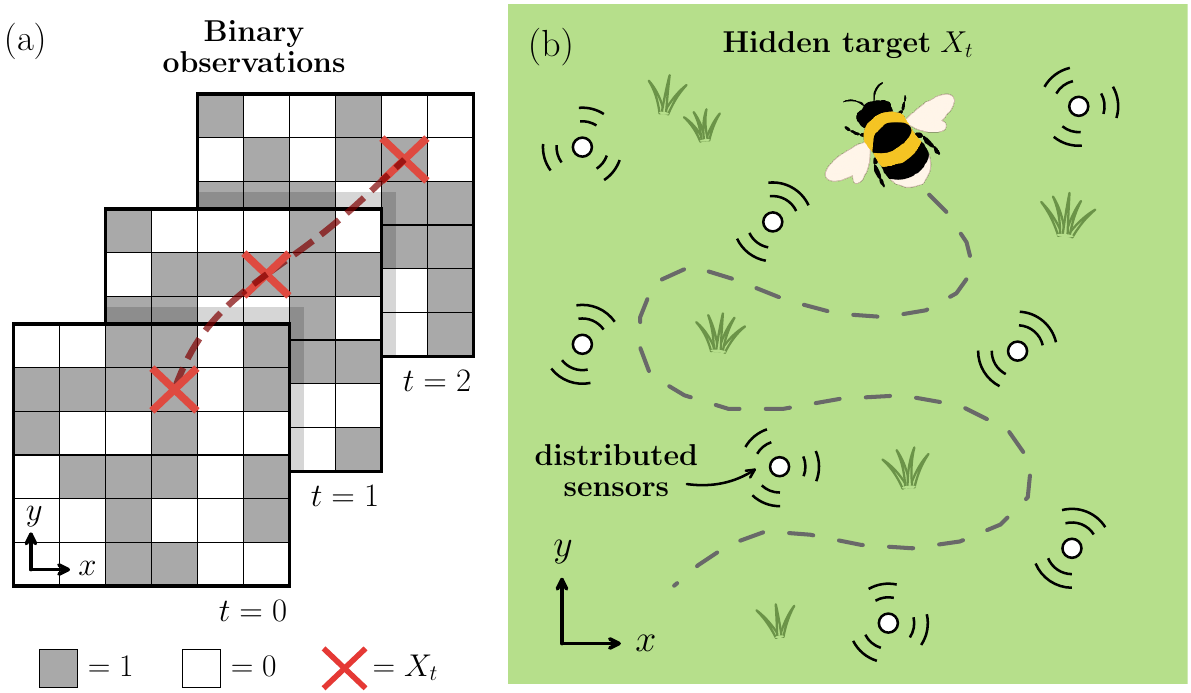}
    \caption{
\textbf{Tracking a hidden target.}
An extended sensor array monitors a diffusing target. At each time, monitoring returns a binary field, with different statistics at the target site than in the background. The observer does not see the trajectory directly, but updates a posterior distribution over candidate positions using Bayes' rule and a dynamical model.
}
    \label{fig:cartoon}
\end{figure}

Remarkably, there is nontrivial universal physics even in the monitoring of a \textit{single} classical random walker \cite{offer2018phase,jin2022kardar,Kim2025,Gerbino2025,yajima2024multifractality,jin2024measurement}. 
Versions of this problem were introduced in Refs.~\cite{offer2018phase,jin2022kardar,Kim2025}.
The task is to infer the hidden position of the particle from the weak signature it leaves in a noisy, spatially extended measurement field (an idealized problem that is reminiscent of more structured real-world tracking or distributed sensing tasks \cite{smal2006bayesian,yuille2002fundamental,frishman2020learning}). The setup is illustrated schematically in Fig.~\ref{fig:cartoon}. Noise means that most of the ``detections'' of the particle are false positives, so it is not obvious a priori whether Bayesian filtering 
(combining information from the noisy measurements with a model for the particle's stochastic motion) will succeed in localizing the particle to a region of a finite characteristic size ${\ell}$, or whether $\ell$ will diverge with the total system volume.  

At Bayes-optimality  (see below \cite{zdeborova2016statistical}) tracking succeeds in ${d=1}$ spatial dimensions \cite{offer2018phase,Kim2025} while there is evidence for a phase transition for $d\geq 3$ \cite{offer2018phase}. One  continuum formulation was discussed in \cite{jin2022kardar}.  Ref.~\cite{Kim2025} gave a formulation related to  the directed polymer (or to the KPZ equation with a pin) 
which we will use extensively here.
The problem has been solved in a mean-field regime (on a tree) \cite{Kim2025,Gerbino2025}. 
But a theoretical understanding of the phase diagram in two or three spatial dimensions has been lacking: we provide this here.

From the point of view of statistical physics, the conditioned ensemble of particle trajectories maps \cite{Kim2025}   to a  partition function for a ``directed polymer in a random environment''~\cite{PhysRevLett.58.2087,PhysRevLett.55.2923,KPZ,huse1985huse,huse1985pinning,HALPINHEALY1995215,Bertini1995} (DPRE).
Imagine a binary occupation measurement ${\eta_t(x)=0,1}$ for each point $(x,t)$ in discrete space and time: noise means that $\eta_t(x)$ can be nonzero even for points $(x,t)$  that do not lie on the target's trajectory. Candidate trajectories 
(worldlines)
are polymer paths: 
the measurement record supplies a  spacetime potential that includes both  homogeneous randomness and an attractive defect along the true trajectory~\cite{Kim2025}.  Equivalently, in the continuum limit, the log-posterior obeys a Kardar--Parisi--Zhang (KPZ) equation with a diffusing attractive defect~\cite{jin2022kardar,Kim2025}. The success or failure of tracking is then the question of whether the defect pins the polymer.
In one dimension the polymer is pinned for any positive signal strength~\cite{tang1993directed,balents1994disorder,hwa1995disorder,LassigKinzelbach1997, Kim2025, jin2022kardar} so that tracking is asymptotically successful. Higher dimensions $d>1$ are different: in analogy to related pinning problems \cite{LassigKinzelbach1997,tang1993directed,Balents_1993,LKL}  
we expect depinning to be possible at finite defect  strength at least in large enough $d$~\cite{offer2018phase,jin2022kardar,Kim2025,Gerbino2025}.

We study tracking in ${d=2,3}$, both at Bayes-optimality---i.e., when the inferrer assumes the correct models for the dynamics and the measurement process---and under a model misspecification,  in which the inferrer makes a false assumption about the target's measurement signature.
Varying  the true and assumed ``noise rates''  (we will be more precise below) gives a two-parameter phase diagram for tracking that includes a Bayes-optimal line. We determine the phase
structure  using  RG at weak coupling (strong noise), stability arguments at strong coupling,
and numerical simulations. 

In ${d=2}$, Bayes-optimal tracking is always asymptotically successful, though with a position uncertainty 
that grows exponentially fast with the noise strength. On the other hand, numerical data at strong coupling  reveal a  transition between a pinned phase and an unpinned phase in \textit{non}-optimal tracking  where the inferrer is over-confident about the measurement data. RG at weak coupling fixes the asymptotic form of the phase boundary, which approaches the Bayes-optimal line tangentially.

In $d=3$ we find a richer phase diagram, with the pinned phase meeting two distinct unpinned phases
at a multicritical point on the Bayes-optimal line. 
The two unpinned phases represent \textit{two different ways} tracking can fail. Underconfidence in the data,  or inadequate data,
can give ``diffuse'' failure, with the posterior spread across the entire system, while overconfidence can give ``localized-but-wrong'' failure, with a posterior that is strongly peaked in the wrong place.  The two phases correspond to Edwards-Wilkinson (EW) and KPZ phases \cite{KPZ,jin2022kardar} for the log-posterior. (Only the latter occurs in ${d=2}$.)
We constrain the shape of the 3+1D phase boundaries using identities that hold on the Bayes-optimal line (and using the simple fact that moving away from Bayes-optimality  cannot improve tracking success).

We characterize the phases and transitions using observables with a direct tracking meaning. For a  target position $X$ and posterior of candidate positions $P(x)$ at a given time $t$, the ``contact fraction'', $P(X)$,  is the posterior weight at the target position  \cite{offer2018phase}.
There are also two useful lengthscales.
The root-mean-square (RMS) distance \cite{Kim2025}
\begin{equation}\label{eq:rmslength}
    \ell_{\rm rms} \equiv \Big(\sum_x P(x)\,|x-X|^2\Big)^{1/2},
\end{equation}
measures how well the posterior is localized on the target, while the width of the posterior distribution is probed by the inverse participation length~\cite{yajima2024multifractality}, defined by ${(\ell_{\rm ipr})^{-d} = \sum_x P(x)^2}$.  (``Localized-but-wrong'' failure is characterized by a large $\ell_{\rm rms}$ and a small $\ell_{\rm ipr}$.) Being more precise, $P(x)$ above could either be the posterior probability at the \textit{final} time---which we will denote simply by $P(x)$---or it could be the distribution at an intermediate time  (computed using measurement data from both earlier and later times). We denote the latter by $\mathcal P(x)$: this will be more useful to us as it allows us to  connect with the polymer pinning literature where $\ell_{\rm rms}$ and  the contact fraction at the pin 
% (computed at intermediate ``times'')
are standard order parameters~\cite{LKL,LassigKinzelbach1997,Toninelli2008,Giacomin2006,Derrida2009,Giacomin2009}.

Finally, we study the universal properties of the various phase transitions in the model. 
Surprisingly, some of the transitions in 3+1D turn out to differ qualitatively from the naive expectation from the $\epsilon$ expansion. 
This connects to interesting and in some cases longstanding  puzzles about critical phenomena for directed polymers in $3+1$D.
We relate the inference problem to recent rigorous developments in the theory of the directed polymer in a random medium~\cite{Junk2022,10.1214/24-AOP1716,Junk2023LocalLT,junk2024strong,junk2024tail,Junk2025coincidence,lacoin2025localization,Zygouras2022,lacoin2025short,zygouras2024directed}. (We also note that the ``size-biased measure'' that appears in various proofs \cite{zygouras2024directed,Birkner2004} has a natural interpretation in terms of Bayesian inference.)

The phase transitions between the ``overconfident failure'' (KPZ) and ``successful'' (pinned) phases 
are in some ways the most straightforward.
We relate them to depinning from a static columnar defect, and compute exponents in 3+1D. (We also derive a scaling relation for the contact-fraction critical exponent at these 
pinning transitions  which appears to be new.)

The other 3+1D transitions (on the Bayes optimal line and in the underconfident regime) 
are more surprising.
We use recent results for the DPRE to argue that they are infinite-order transitions---as the usual DPRE temperature-driven transition was recently shown to be~\cite{lacoin2025localization} and as some other pinning transitions have been argued to be \cite{tang2001rare,derrida2014depinning,monthus2017strong,chen2021derrida}, but in contrast to what would naively be expected from the $\epsilon$ expansion.
We make some preliminary remarks
(relevant also to the conventional DPRE and surface roughening transitions)
about how such infinite order transitions can emerge from the RG.

A remaining puzzle is that, over numerically accessible sizes, the data for these (putatively infinite-order) inference transitions remain well described by conventional finite-size scaling, with apparently finite effective exponents. Resolving this tension, for example by understanding finite-size effects at the infinite-order transitions,
remains an interesting open problem.

\vspace{4mm}
\tableofcontents

\section{Tracking model}
\label{sec:model}

We consider a target diffusing on a $d$-dimensional hypercubic lattice. 
At each discrete time step it hops to one of its $2d$ nearest neighbors, each with probability ${2d \times D}$, where $D$ is the diffusion constant, and otherwise remains in place. This defines the hopping kernel
\begin{equation}
K(x|y)
=
(1-2dD)\,\delta_{x,y}
+
D{\sum}_{|e|=1}\delta_{x,y+e}.
\label{eq:hop_kernel}
\end{equation}
For concreteness we take the lattice to be an ${L\times \cdots \times L}$ torus.

Let the true position of the target at time $t$ be $X_t$.
At every timestep, each site yields an independent binary observation $\eta_t(x)\in\{0,1\}$.   Let $p_0$ and $p_1$ denote the probability of observing $\eta_{t}(x)=1$ in the case where the site is empty (${X_t\neq x}$) and when it is occupied (${X_t=x}$), respectively. That is, $p_0$ is the probability of a false positive, and ${1-p_1}$ is the probability of a false negative; noiseless measurement gives ${p_0=0}$, ${p_1=1}$. So long as ${p_1>p_0}$, the measurement signal is stronger on average at the true location $X_t$ than at other locations:
\begin{align}
{\mathbb{E}\, [\eta_t(x) | X_{t}]}
= p_0 + (p_1-p_0) \, \delta_{x,X_t}.
\label{eq:measurementmean}
\end{align}
Here $\mathbb{E} [\bullet|X_t]$ is the average over $\eta$ at fixed $X_t$. (See Refs.~\cite{offer2018phase,Kim2025,jin2022kardar,Gerbino2025} for other measurement protocols.)

Bayes optimal inference is the case where the inferrer knows the values of $p_0$ and $p_1$, as well as the hopping kernel $K$. We also consider the non-optimal case where the inferrer assumes incorrect values $p_0'$ and $p_1'$. 
For the majority of the paper we will assume that the inferrer knows the true  hopping kernel $K$,  and therefore the true value of the diffusion constant $D$. However, once this case is understood, many  of the results extend directly to the case where the inferrer assumes an incorrect diffusion constant $D'$, and we will comment on this towards the end of the paper (Sec.~\ref{sec:misspecificationofD}).

To formalize the Bayesian ``filtering'' (i.e. the inference of the instantaneous state; we will consider the full trajectory in a moment) it is useful to introduce the log-odds contrasts 
for the true and assumed models:
\begin{equation}
\Delta
\equiv
\ln \!\left(\frac{p_1/(1-p_1)}{p_0/(1-p_0)}\right),
\quad
\Delta'
\equiv
\ln \!\left(\frac{p_1'/(1-p_1')}{p_0'/(1-p_0')}\right).
\label{eq:contrasts}
\end{equation}
Larger $\Delta$ means better measurement data.
Writing  ${P_t(x)=\Pr'\big(\hspace{-0.1mm}{X_t=x}\hspace{-0.4mm}\mid\hspace{-0.4mm} \eta_{1:t}\hspace{-0.1mm}\big)}$
for the filtered posterior distribution for the instantaneous target location, under the assumed model,
Bayes' theorem gives
\[
P_{t+1}(x)\propto {\rm Pr}' \Big( \eta_{t+1}
\hspace{-0.3mm} \mid \hspace{-0.3mm}
X_{t+1}=x\Big) {\sum}_{y}K(x|y)\,P_t(y),
\]
where ${\eta_t = \{ \eta_t (y)\}_y}$.
The first factor on the right-hand-side is the probability of the measurement ``snapshot'' $\eta_{t+1}$, according to the assumed model, conditioned on a given target location.
By the conditional independence of the sitewise observations, this term is a product of $p_0'$ and $p_1'$ factors, and reduces simply to $e^{\Delta'\eta_{t+1}(x)}$ up to an $x$ independent factor.
So, writing ${P_t(x)\propto Z_t(x)}$, 
where $Z_t(x)$ is an unnormalized posterior, gives the update rule
\begin{equation}
Z_{t+1}(x)
=
e^{\Delta' \eta_{t+1}(x)}
{\sum}_{y}
K(x|y)
\,Z_t(y)
\label{eq:Zupdate}
\end{equation}
for the posterior.
We see that $\Delta'$ quantifies the confidence that the inferrer has in the measurement data (while $\Delta$ is the  confidence that they ``ought'' to have in the data).

Next, consider spacetime trajectories. We choose the  initial condition ${X_0=0}$, ${P_0(x)=\delta_{x,0}}$. In a finite system, the initial condition becomes irrelevant at late times ${t/L^2\gg 1}$.\footnote{Note that, depending on the prior, the behavior may differ for the order of limits $\lim_{L\to \infty}\lim_{t\to\infty}$ and for the order $\lim_{t\to \infty}\lim_{L\to\infty}$. For example, if the prior is flat, and we take $L\to \infty$ at finite $t$ then at finite $\Delta'$ we have no chance of finding the particle.}

Iterating Eq.~\ref{eq:Zupdate} with this initial condition
gives  $Z_t(x)$ as a sum over ``candidate''  trajectories~\cite{Kim2025}:
\begin{align}
Z_t(x) = \sum_{x_{t-1},\ldots, x_1}
 e^{\Delta'\sum_{\tau=1}^t \eta_\tau(x_\tau)}
\prod_{\tau=1}^{t}
K(x_\tau|x_{\tau-1}),
\label{eq:worldinedisorder}
\end{align}
with ${x_0=0}$ and ${x_t=x}$.
This ``Boltzmann weight''---the summand---also defines the Bayesian posterior probability for full spacetime trajectories, conditioned on the full measurement record.

Let us imagine fixing both the true trajectory $\{X_\tau\}$ and the measurement record $\{\eta_\tau(x)\}$.
Eq.~\ref{eq:worldinedisorder} then defines 
a partition function for a polymer in a random potential $\eta_t(x)$.
The polymer is the candidate trajectory
$\{x_\tau\}$; as a   result of (\ref{eq:measurementmean}), the polymer is effectively attracted to the true trajectory $\{X_\tau\}$.
In this point of view, we think of both $\{X_\tau\}$ and $\{\eta_\tau(x)\}$ as the   ``quenched disorder'' defining the polymer partition function.\footnote{In our specific microscopic model, 
labelling the partition function by both $\{X_\tau\}$ and $\{\eta_\tau(x)\}$ is strictly speaking redundant, since $Z_t(x)$ depends directly only on $\{\eta_\tau(x)\}$. 
But this redundant labelling is convenient: in view of Eq.~\ref{eq:measurementmean},
it makes sense, after a bit of coarse-graining, to talk of the polymer being attracted to the true trajectory. This is apparent in the  continuum descriptions below.}
We will often refer to $\{X_\tau\}$ as an attractive ``pin'' for the polymer.

For future use, we note that
Eq.~\ref{eq:worldinedisorder} also gives the posterior
distribution at an intermediate time. Consider a record of total duration \(t\), and
let \({0<\tau<t}\). Marginalizing the posterior trajectory ensemble over all
time slices except \(\tau\) gives the posterior~\cite{Kitagawa1994}
\begin{equation}
    \mathcal P_{\tau}(x)
    \equiv
    {\rm Pr}'\!\left(X_\tau=x\,\middle|\,\eta_{1:t}\right)
    =
    \frac{
        Z_\tau^{\rightarrow}(x)Z^{\leftarrow}_{t-\tau}(x)
    }{
        \sum_y Z_\tau^{\rightarrow}(y)Z^{\leftarrow}_{t-\tau}(y)
    } .
    \label{eq:two_sided_posterior}
\end{equation}
Here \(Z_\tau^{\rightarrow}(x)\) is the ``forward'' partition sum (for the  first part of the trajectory),
which can be obtained 
from real-time tracking with Eq.~\ref{eq:Zupdate} using the record \(\eta_{1:\tau}\). Since the random walk kernel is symmetric, 
the partition sum \(Z^{\leftarrow}_{t-\tau}(x)\) for the second part of the trajectory
may be obtained from the real-time tracking formula (\ref{eq:Zupdate}) with a reversed measurement record \(\eta_{t:\tau+1}\).\footnote{Since ${Z^\leftarrow_{0}=1}$, the  ``initial'' condition for the backward tracking problem above is a flat ($x$-independent) distribution, which  differs from the forward tracking problem. These boundary conditions become unimportant for large enough $\tau$ and $t-\tau$ in a finite system. In  numerical simulations we will use a flat prior,  which leads to more symmetric boundary conditions (App.~\ref{app:numerics})}.
Eq.~\ref{eq:two_sided_posterior} is thus simply the  normalized 
product of posteriors obtained by forward and backward  real-time filtering. 
We refer to this inference for a past state 
(sometimes referred to as ``smoothing'' \cite{Sarkka_2013}) as retrospective tracking.

\section{Continuum descriptions}
\label{sec:ctmdescr}

Writing ${h= \ln Z}$, Eq.~\ref{eq:Zupdate} is a discrete version of a KPZ equation with a defect tied to the true trajectory~\cite{Kim2025} (see also a related formulation \cite{jin2022kardar}). 
This continuum description can be obtained in a controlled way in the present model whenever $\Delta'$ is small, but captures universal properties of the transitions more generally:
\begin{equation}
\partial_t h
=
D\nabla^2 h + D(\nabla h)^2 + \xi(x,t)
+
\mathsf g_{st}\,\delta^{(d)}(x-X_t),
\label{eq:kpz_defect}
\end{equation}
with $\xi$ Gaussian white noise of variance $\mathsf g_{ss}$.
General expressions for $\mathsf g_{ss}$ and $\mathsf g_{st}$ are given  in the Supp. Mat.~\cite{suppmat}; they 
 take a simple form when ${p_0=p_0'=\tfrac12}$,
which we will use for most of our simulations:
\begin{align}
\mathsf g_{ss}&=\frac{(\Delta')^2}{4},
&
\mathsf g_{st}&=\frac{\Delta\Delta'}{4}.
\label{eq:couplings_from_Delta}
\end{align}
We take Eq.~\ref{eq:couplings_from_Delta} as definitions of $\mathsf g_{ss,st}$ even when $\Delta, \Delta'$ are not small.
The Nishimori condition 
for Bayes-optimality
is ${\mathsf g_{st}=\mathsf g_{ss}}$. 
 Overconfident inference is  ${\mathsf g_{ss}>\mathsf g_{st}}$ 
and underconfidence is 
${\mathsf g_{ss}<\mathsf g_{st}}$.

The dimensionful parameter $D$ can be scaled out of Eq.~\ref{eq:kpz_defect}, leaving only \textit{two} dimensionless coupling constants (see below).
That is, despite the various microscopic parameter choices involved in defining the model, 
the phase diagram is essentially controlled by two parameters (and by only one in the Bayes-optimal case).

An equivalent continuum formulation is based
on applying the replica trick to the spacetime picture  in Eq.~\ref{eq:worldinedisorder}
above. Initially, the coupling $\mathsf g_{st}$
measures the attraction of the candidate polymer  to the true trajectory, while $\mathsf g_{ss}$ sets the variance of the bulk disorder.
After performing the disorder average over $\eta$ using the replica trick,
the candidate polymer $x_\tau$ is replaced with ${n-1}$ replicas $x_{\tau, a}$ for ${a=1,\ldots, n-1}$ which are referred to as ``students'' \cite{zdeborova2016statistical}.
The  unusual feature here is the 
presence of an additional 
``teacher''
polymer ${x_{\tau,0}=X_\tau}$
representing the true trajectory:
\begin{equation}
S_n
=
\sum_{a=0}^{n-1}\int dt\,{\frac{\dot x_a^2}{4D}}
-
\sum_{a<b}\mathsf g_{ab}\int dt\,
\delta^{(d)}\bigl(x_a(t)-x_b(t)\bigr).
\label{eq:replica_polymer_action}
\end{equation}
The interaction is \(\mathsf g_{ab}=\mathsf g_{st}\) for \(a=0\), i.e. for teacher-student pairs,
and \(\mathsf g_{ab}=\mathsf g_{ss}\) otherwise, i.e. for student-student pairs (UV regularization of the interaction at momentum ${\Lambda\sim 2\pi}$ is implied \cite{suppmat}).
Ultimately the replica limit ${n\to 1}$ must be taken.

The above formulation can be applied for a fixed $\{X_\tau\}$, in which case the ${a=0}$ term in the first sum plays no role. 
But usually we wish also to average over $\{X_\tau\}$ in order to obtain results for typical instances of the tracking problem. For this reason we have included the correct diffusive continuum weight for the teacher polymer as well as the students.

In the present model the assumed and true target dynamics are the same, i.e. $D$ is the same for teacher and students (see Sec.~\ref{sec:misspecificationofD} for the more general case).
The  \(S_{n-1}\) permutational symmetry among the students is therefore enhanced to full \(S_n\) symmetry at Bayes optimality.
Eq.~\ref{eq:replica_polymer_action} is also equivalent to a theory of bosons with contact interactions~\cite{PhysRevLett.55.2235,KARDAR1987582,dotsenko2010replica,calabrese2011exact,nakayama2021efimov}.

\section{Weak-coupling RG}
\label{sec:weakcoupling}

Power counting for Eq.~\ref{eq:replica_polymer_action} gives the critical dimension \(d_c=2\) in which weak interactions are marginal. 
Because the  interaction vertex preserves the labels on the worldlines, each of the ${n(n-1)/2}$ couplings $\mathsf g_{ab}$ renormalizes independently of all others within perturbation theory.
The beta functions for the dimensionless couplings
$g_{ab}=\frac{K_d}{2D}\Lambda^{d-2}\mathsf g_{ab}$
[with \(K_2=1/(2\pi)\) and \(K_3=1/(2\pi^2)\)]
are, to all orders,
\begin{equation}
{d g_{ab}}/{dl}
=
(2-d) g_{ab}+ g_{ab}^{\,2},
\label{eq:generic_beta}
\end{equation}
where \(l\) is the logarithm of the spatial rescaling factor. (See Appendix~\ref{app:replica-boson-RG} for a derivation.) The autonomous flow for each $g_{ab}$ means Eq.~\ref{eq:generic_beta} can in fact be inferred from the standard KPZ case \cite{Wiese1998}.

The key utility of this beta function is for analyzing the stability of the fixed point at ${g_{ab}=0}$. Together with stability arguments at strong coupling, this will  allow us to determine the universal features of the phase diagram in ${d=2}$, and the topology of the phase diagram in ${d>2}$.
An important caveat is that we do not expect perturbative RG (even to all orders~\cite{Wiese1998}) to correctly predict the critical properties of the nonzero fixed points in $d=3$. As discussed in Sec.~\ref{sec:infinite-order-transitions}, recent rigorous results show that the bulk DPRE roughening transition is infinite order and is asymptotically governed by the free fixed point, rather than by the finite-coupling fixed point predicted by the $\epsilon$ expansion.

\begin{figure*}[!t]
\includegraphics[width=1.0\linewidth]{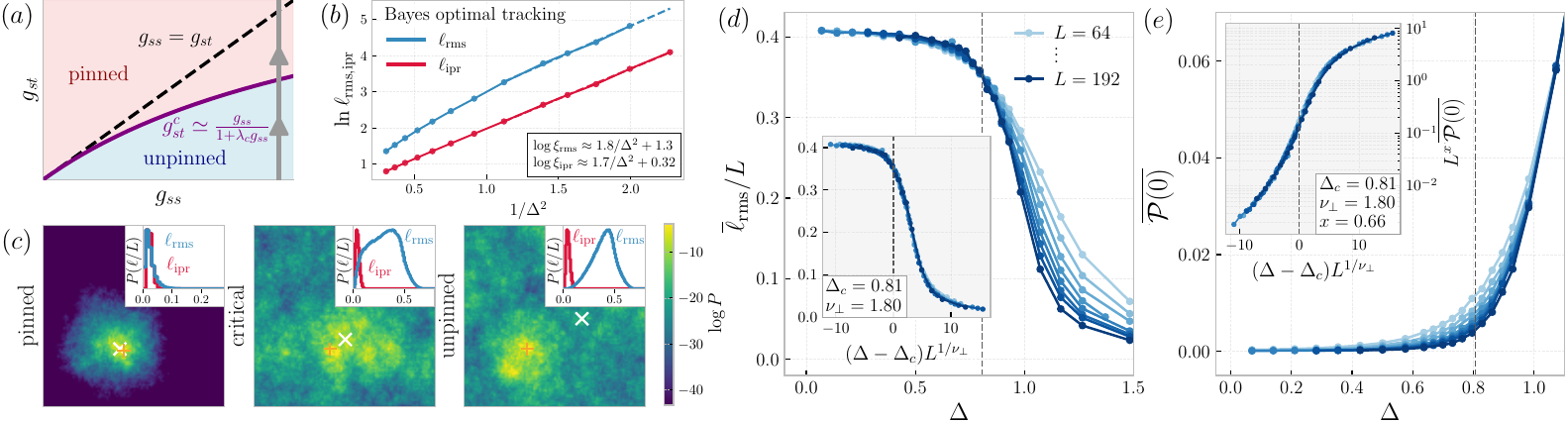} 
\caption{
\textbf{Tracking \(d=2\) dimensions.}
{\bf (a)}~Schematic phase diagram: the depinning boundary approaches the Bayes optimal line \(g_{st}=g_{ss}\) tangentially at weak coupling. The gray line represents the scan used in panels (d,e).
{\bf (b)}~The RMS length \(\ell_{\rm rms}\) and inverse-participation length \(\ell_{\rm ipr}\equiv(\sum_x {\mathcal P}(x)^2)^{-1/2}\) as a function of measurement strength at Bayes optimality $g_{ss}=g_{st}=\Delta^2/4$, showing essential growth \(\ell\sim\exp(A/\Delta^2)\); here \((D,p_0,L) = (0.0375,0.5,192)\) with \(N_{\rm samp}^{\rm Nish}\simeq 10^7\) 
{\bf (c)}~Posterior snapshots in the pinned, critical, and unpinned regimes, taken at true measurement contrast $\Delta = 1.73, 0.86, 0.28$ respectively [varying $p_1$ with \((D,p_0,p_0',p_1')=(0.05,0.5,0.5,0.85)\) fixed. All snapshots use $L=96$]. The true target location is shown with the white `$\times$', and the 
posterior maximum with the orange `$+$'. Insets show distributions of \(\ell_{\rm rms}\) and \(\ell_{\rm ipr}\).
{\bf (d,~e)}~Finite-size scaling of the localization length \(\ell_{\rm rms}/L\) and contact fraction \({\overline{\mathcal P(0)}}\).
We vary the true measurement model at fixed assumed model (varying $\Delta$). The insets show the corresponding collapses. Parameters are \((D,p_0,p_0',p_1')=(0.05,0.5,0.5,0.85)\) (corresponding to $\Delta'\simeq 1.73$, while $\Delta$ is varied via $p_1$) with \(L=40,\ldots,196\). 
Data are gathered after time \(10L^2\) to reach stationarity, and use \(N_{\rm samp}\simeq 2\times 10^6\) samples for each $(L,\Delta)$.}
\label{fig:2D_main}
\end{figure*}

\section{Tracking  in ${d=2}$ spatial dimensions}
\label{sec:tracking2D}

To diagnose tracking, we will use lengthscales \(\ell_{\rm rms}\) \eqref{eq:rmslength} and $\ell_{\rm ipr}$, and the contact fraction, written \(\mathcal P(0)\) in the teacher-centred frame. We define both $\ell_{\rm rms}$ and $\ell_{\rm ipr}$  using the posterior $\mathcal P(x)$ of the retrospective tracking problem (Eq.~\ref{eq:two_sided_posterior}),
but the final-time observables show qualitatively similar behavior.

In a pinned phase, \(\overline{\ell}_{\rm rms}/L\) and \(\overline{\ell}_{\rm ipr}/L\) vanish as ${L\to\infty}$, and \(\overline{\mathcal P(0)}=O(1)\). This is a phase where inference ``succeeds''.
In an unpinned phase, 
the posterior detaches from the truth, 
so \(\overline{\ell}_{\rm rms}/L=O(1)\) and \(\overline{\mathcal P(0)}\to0\). 
As we will discuss, different kinds of unpinned phase are possible and are distinguished by the behavior of $\overline{\ell}_{\rm ipr}/L$.

At a depinning critical point, scale invariance implies that \(\overline{\ell}_{\rm rms}/L\)
and 
\(\overline{\ell}_{\rm ipr}/L\)
approach nonzero universal values,
while the averaged contact fraction vanishes algebraically, $\overline{\mathcal P(0)}\sim L^{-x}$, defining the contact-fraction exponent \(x\).\footnote{For real-time (as opposed to retrospective) tracking we expect  
$\overline{P(0)}\sim L^{-x'}$ with 
$x'\leq x$.} Here $\overline{(\bullet)}$ denotes an average over target trajectories and measurement records at late time, or equivalently, a long-time average in the simulations.

\subsection{Phase diagram in 2D}
\label{subsec:2Dphasediag}

The topology of the 2D phase diagram is shown in Fig.~\ref{fig:2D_main}~(a), and can be argued for as follows. 

First, note that, for any $g_{ss}$, the system is in a stable \textit{pinned} phase at least for large enough  $g_{st}$. 
This is apparent in the  language of the polymer in a disordered environment with an attraction to the true trajectory $\{X_\tau\}$:
for large $g_{st}$, the pinning energy is large  and the polymer must remain close to the true trajectory in order to optimize its free-energy-per-unit-length. In this phase, the posterior $P(x)$ typically decays exponentially away from the true location (because of the extensive free energy cost of long excursions away from the pin).

This stability of the pinned phase at large enough $g_{st}$ holds in any $d$; 
in ${d=2}$, as we show below, the pinned phase includes the entirety of the Bayes-optimal line.
Fig.~\ref{fig:2D_main} (c) shows a snapshot of the final-time posterior $P(x)$ within the pinned phase (specifically, at Bayes optimality). The inset shows the histogram of $\ell_{\rm rms}/L$ and $\ell_{\rm ipr}/L$, consistent with the expected localization.

Second, there is a stable \textit{unpinned} phase in the sufficiently overconfident regime.
Consider the limit
\({g_{st}=0}\), 
where the candidate polymer sees only uncorrelated spacetime randomness and has no attraction to the true  trajectory.
This limit is therefore the standard 
\(2+1\)D DPRE, and the log-posterior $h$ is governed by the KPZ equation.
The resulting KPZ phase is \textit{stable} to weak \(g_{st}\). 
To see this, first note that 
 the diffusive wandering of the teacher is irrelevant at the KPZ fixed point, because the 
dynamical exponent $z=2$ of the diffusive wandering is larger than the \(2+1\)D KPZ dynamical exponent ${z_\text{KPZ}\simeq 1.6}$.
(Recall that $z_\text{KPZ}$ determines the appropriate relative rescaling of space and time when we perform RG.)
A weak teacher signal therefore coarse-grains to the columnar (static) defect problem, and the KPZ fixed point is stable against a  weak columnar defect when ${d>1}$~\cite{tang1993directed,balents1994disorder,hwa1995disorder,LassigKinzelbach1997}. 

Thus overconfidence allows a stable failure phase.  In this phase the posterior is sharply localized at late times, but typically at the wrong position. Localization is due to the fact that the KPZ height field $h(x,t)$ 
varies by an amount of order ${\ell^{\chi_\text{KPZ} } \gg 1}$ for points separated by a distance $\ell$ \cite{KPZ} (with ${\chi_\text{KPZ}= 2-z_\text{KPZ}\simeq 0.4}$ \cite{pagnani2015numerical}).
This scaling implies orders-of-magnitude differences between  posterior probabilities $P(x)\propto e^{h(x)}$ for well-separated points,\footnote{Since the posterior at intermediate times is given, up to normalization, by the product of two independent real-time-tracking posteriors, ${\mathcal P(x) \propto P^{\rightarrow}(x) P^{\leftarrow}(x) \propto e^{h^{\rightarrow}(x) + h^{\leftarrow}(x)}}$, the same conclusion applies for retrospective tracking.}
and an ${\ell_{\rm ipr}}$ that is typically of order 1.
Therefore we expect that ${\ell_{\rm ipr}/L\to 0}$ in this phase, despite the fact that ${\ell_{\rm rms}/L=\mathcal{O}(1)}$.   
A representative  snapshot of the posterior is shown in  Fig.~\ref{fig:2D_main} (c); the histograms in the inset are consistent with our expectations for  $\ell_{\rm rms,ipr}/L$.

Having established the existence of two phases, next we argue that  Bayes-optimal tracking is pinned for any nonzero measurement contrast $\Delta$,
i.e. that in 2+1D the diagonal ${g_{st}=g_{ss}=g}$ of the phase diagram lies within the pinned phase.
This line is preserved under RG, and Eq.~\ref{eq:generic_beta} shows that it flows to larger $g$, i.e. to larger $\Delta$. 
The natural hypothesis is that this flow leads to the stable pinned phase. 
Combined with this hypothesis, the marginal flow ${\dot g = g^2}$ leads to the prediction that the localization lengths grow exponentially at small $\Delta$  (see Supplemental Material):
\begin{equation}
 \ell_{\rm rms,ipr}
 \asymp
\exp \left( \frac{16\pi D}{\Delta^2} \right).
\end{equation}
The simulation results in Fig.~\ref{fig:2D_main} (b) indeed agree well with this exponential scaling, 
confirming the hypothesis.
The figure shows a case with
\(16\pi D\simeq1.89\);  the measured slopes are $\Delta^2 \log \ell_{\rm rms/ipr}\approx 1.7\text{--}1.8$. 

This scaling also implies that $P(0)$ is exponentially small at small $g$, explaining the prescient  observation in Ref.~\cite{offer2018phase} that  $P(0)$ appeared to decrease exponentially with ``clutter'' density in simulations of an alternative lattice model for particle tracking.

Weak-coupling RG also gives a prediction for the shape of the depinning phase boundary at small ${(g_{ss}, g_{st})}$. Eq.~\ref{eq:generic_beta} gives
${\partial_\tau(g_{st}^{-1}-g_{ss}^{-1})=0}$,
showing that   flow lines can be labelled by \(g_{st}^{-1}-g_{ss}^{-1}\). 
Since the phase boundary is preserved under RG, it must coincide with a flow line, i.e.  the phase boundary is  
\(g_{st}^{c} \simeq  g_{ss}/(1+\lambda_c g_{ss})\), for some \({\lambda_c>0}\). 
It follows that the boundary approaches the Bayes-optimal line tangentially as \(g_{ss}\to0\), as illustrated in Fig.~\ref{fig:2D_main} (a).

\subsection{Critical properties  in ${d=2}$}
\label{sec:crit2d}

It remains to discuss the depinning phase transition itself.
The teacher's diffusion is in fact again irrelevant at this transition, just as it was in the unpinned phase. 
At criticality, long excursions of the polymer
away from the defect (teacher) are possible,
and these excursions necessarily obey standard DPRE (KPZ) scaling with ${z_{\rm KPZ}\simeq 1.6}$ as above. 
Therefore 
${z_{\rm KPZ}}$ is also the only consistent possibility for the dynamical exponent at the critical fixed point (and determines the appropriate RG transformation).
Since ${z_{\rm KPZ}<2}$,
the diffusively wandering teacher again coarse-grains to a static
``columnar'' defect.
Therefore we expect the transition to be in the universality class for depinning of a DPRE from a columnar defect in $2+1$D, which was studied in Ref.~\cite{LKL}.

We study the transition numerically by reducing the true measurement contrast \(\Delta\) at fixed assumed contrast $\Delta'$,  giving a vertical scan in the \((g_{ss},g_{st})\) plane. 
We set $D=0.05$ and \(\Delta'=1.73\) (we do not expect the choice of values to affect the universal physics).
% \(p_{\rm hop}=0.15\). 

Scale invariance at the transition implies that the dimensionless quantity \(\overline{\ell}_{\rm rms}/L\) takes a universal value there. Indeed, plotting  \(\overline{\ell}_{\rm rms}/L\) versus $\Delta$,
we find that the curves for 
different $L$ cross, locating the  depinning transition at \(\Delta_c\simeq0.81\). 
We also obtain good scaling collapses of \(\overline{\ell}_{\rm rms}/L\) and \(\overline{\mathcal {P}(0)}\)
using  the static-defect critical exponents of Ref.~\cite{LKL}, \(\nu_\perp\simeq1.80\) and \(x\simeq0.66\).
The  corresponding scaling forms are ${\overline{{\mathcal P}(0)}=L^{-x} F[ (\Delta-\Delta_c)L^{1/\nu_\perp}]}$
and 
${\overline{\ell}_{\rm rms}/L =H[(\Delta-\Delta_c)L^{1/\nu_\perp}]}$, for some scaling functions $F$ and $H$.
Here the correlation length exponent $\nu_\perp$ also sets the size of the ``bound state'' just inside the pinned phase, ${\overline{\ell}_{\rm rms} \sim (\Delta-\Delta_c)^{-\nu_\perp}}$.

These collapses are shown in the insets to  Figs.~\ref{fig:2D_main} (d,e), 
and support the above characterization of the fixed point.
As argued in Sec.~\ref{sec:contactfrac}, $x$ is not in fact an independent exponent, but is determined by $\nu_\perp$ together with standard KPZ exponents: ${x=2(z_{\rm KPZ}-1) - 1/\nu_\perp}$.
As mentioned above, the dynamical exponent (which relates the correlation length exponent $\nu_\perp$ and the correlation time exponent $\nu_\parallel$ via ${z=\nu_\parallel/\nu_\perp}$)
is equal to the KPZ value, ${z=z_{\rm KPZ}}$.

\begin{figure*} \centering 
\includegraphics[width=1.0\linewidth]{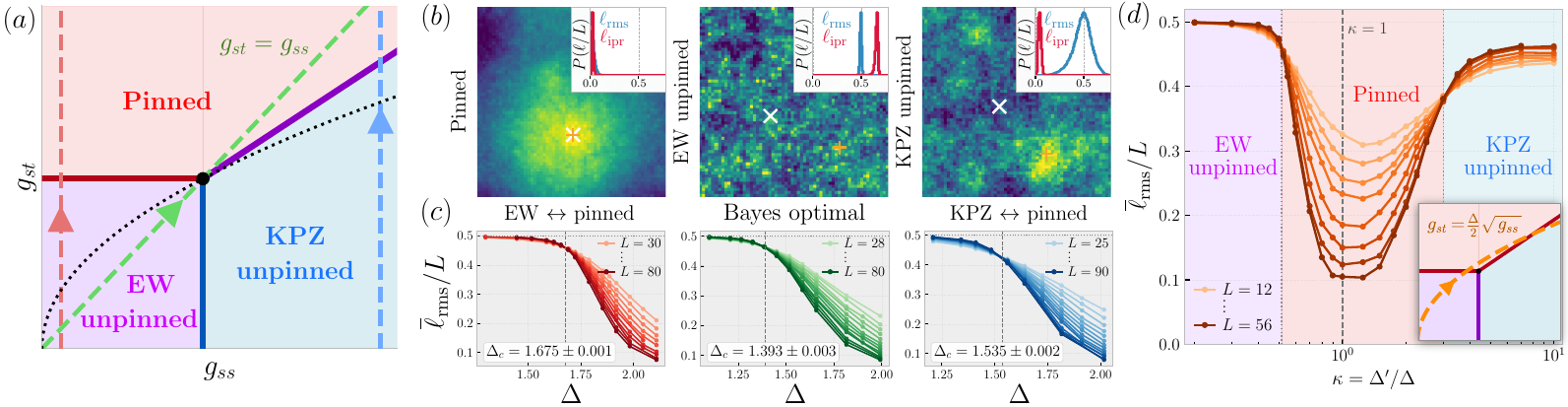} 
\caption{
\textbf{Tracking in \(d=3\) dimensions.}
{\bf (a)}~Schematic phase diagram in the \((g_{ss},g_{st})\) plane.
The pinned phase meets two unpinned phases, EW-smooth and KPZ-rough, at the multicritical point \((g_*,g_*)\) on the Bayes-optimal line \(g_{ss}=g_{st}\).
The red, green, and blue dashed lines indicate the three diagnostic scans shown in panel~(c), while the dotted black curve is the misspecification path through the multicritical point.
{\bf (b)}~Representative posterior snapshots in the pinned
(\(p_0,p_1,p_1'=0.5,0.9,0.9\)), EW-unpinned
(\(p_0,p_1,p_1'=0.1,0.289,0.214\)), and KPZ-unpinned
(\(p_0,p_1,p_1'=0.5,0.725,0.95\)) phases. All snapshots use \(D=0.033\) and \(p_0'=p_0\).
The white \(\times\) marks the true target position and the orange \(+\) the maximum of the posterior. Insets show the distributions of \(\ell_{\rm rms}\) and \(\ell_{\rm ipr}\equiv(\sum_x {\mathcal P}(x)^2)^{-1/3}\).
{\bf (c)}~Finite-size scans of the retrospective tracking RMS distance \(\overline{\ell_{\rm rms}}/L\).
For the EW--to--pinned (red) and KPZ--to--pinned (blue) transitions we use vertical paths through the $(g_{ss},g_{st})$ 
plane,
implemented by varying the true record at fixed assumed model---\((p'_{0},p'_{1})=(0.5,0.95)\) for the KPZ scan and \((p'_{0},p'_{1})=(0.10,0.214)\) for the EW scan. For the Bayes optimal depinning transition (green) we vary $\Delta=\Delta'$ together (varying $p_1$ with fixed $p_0=0.5$). All three scans take \(D = 0.033\); further numerical details are given in Appendix~\ref{app:numerics}. 
{\bf (d)}~Misspecification scan:
fixing the true model, we vary the observer's assumed measurement strength by \(\Delta'=\kappa\Delta\). 
(orange curve in inset).
True model parameters are \((D,p_0,p_1)=(0.0167,0.5,0.76)\), using \({N_{\rm samp}}\simeq 2\times 10^5\) samples for each $(L,\Delta')$. Underconfidence (overconfidence) drives the EW-unpinned (KPZ-unpinned) failure mode. (In all scans we take $p_0'=p_0$.)
}
\label{fig:3D_main}
\end{figure*}

\section{Tracking phase diagram in \({d=3}\)}
\label{sec:tracking3D}

{
Three spatial dimensions are qualitatively different because the Gaussian fixed point at ${g_{ss}=g_{st}=0}$ is stable. This allows an additional phase, representing a different kind of tracking failure from that found in 2D, and allows a finite-coupling transition in Bayes-optimal tracking.

In ${2+\epsilon}$ dimensions, Eq.~\ref{eq:generic_beta} gives $\dot g_{st}=-\epsilon g_{st}+g_{st}^2$ and $\dot g_{ss}=-\epsilon g_{ss}+g_{ss}^2$ at weak coupling. 
These flows suggest the phase-diagram topology shown in Fig.~\ref{fig:3D_main}~(a), with three stable phases: a pinned phase where inference succeeds, and two unpinned phases where inference fails. Within the $\epsilon$ expansion these phases are separated by unstable fixed points at $(g_{ss},g_{st})=(0,\epsilon)$ and $(\epsilon,0)$, together with a doubly-unstable (multicritical) fixed point at $(\epsilon,\epsilon)$.

The nontrivial fixed points of this weak-coupling RG should not, however, be expected to give a quantitatively reliable description in $d=3$. We therefore use the $\epsilon$ expansion only as a 
quick way to motivate the phase-diagram topology:
 the existence of the three phases and the bulk roughening boundary, as well as the fact that the multicritical point lies on the Bayes-optimal line, can subsequently be confirmed by more general arguments.

\subsection{Phase-diagram topology}
\label{sec:3dtopology}

The stable Gaussian fixed point at ${(g_{ss},g_{st})=(0,0)}$ governs an ``EW--unpinned'' phase in which the height field (log posterior) has Edwards--Wilkinson statistics~\cite{barabasi1995fractal}.\footnote{In the stationary state, the height field resembles a free field.} Such a height field is sometimes referred to as ``smooth'', because above two dimensions the variance of the local height fluctuations remains finite in the thermodynamic limit. As a result, the posterior probability is spread broadly across the system: the probability for the particle to lie in a region of volume ${V\gg1}$ is $V/L^3$, up to subleading fluctuations. The middle panel of Fig.~\ref{fig:3D_main}(b) shows a representative posterior in this phase; both $\ell_{\rm rms}$ and $\ell_{\rm ipr}$ are $O(L)$.

There is also a stable pinned phase. As argued in Sec.~\ref{sec:tracking2D}, the strong-coupling pinned phase is stable in any spatial dimension, so for sufficiently large $g_{st}$ the candidate polymer remains bound to the teacher. The posterior is then exponentially localized on the truth, with $\ell_{\rm rms},\ell_{\rm ipr}=O(1)$. A representative posterior is shown in the left-hand panel of Fig.~\ref{fig:3D_main}(b). The boundary between the EW-unpinned and pinned phases is shown by the dark-red line in Fig.~\ref{fig:3D_main}(a).

Finally, there is a stable KPZ-unpinned phase. At ${g_{st}=0}$ the teacher decouples completely, leaving the ordinary $3+1$D directed polymer in a random environment. For sufficiently large $g_{ss}$ this system is in its strong-disorder, KPZ-rough phase. This phase remains stable to sufficiently weak attraction to the teacher: since $z_{\rm KPZ}<2$, the diffusive motion of the teacher is asymptotically irrelevant, and at long scales the problem reduces to a directed polymer in a random medium with a static columnar defect, for which a finite attraction is required to bind the polymer in $d>1$~\cite{LassigKinzelbach1997}. In this phase large free-energy fluctuations sharply localize the posterior, but typically at the wrong position, so that $\ell_{\rm rms}=O(L)$ while $\ell_{\rm ipr}=O(1)$. A representative posterior is shown in the right-hand panel of Fig.~\ref{fig:3D_main}(b).

These three phases can also be distinguished by the overlap 
between student polymers \cite{derrida1988polymers, cook1989polymers, derrida1990directed,Kim2025}, or between teacher and student. 
Let \(q_{\rm ss}\)
be the mean overlap between two posterior samples, written in the replica language \(q_{\rm ss}= t^{-1} \langle \sum_{\tau=1}^t \delta_{x_a(\tau), x_b(\tau)} \rangle 
\) for any $a>b>0$.
Let  \(q_{\rm st}\) be the mean overlap with the teacher: this is 
written similarly but with $a>b=0$,
and is also the time-averaged contact fraction.
In the EW phase,  \(q_{\rm ss}=q_{\rm st}=0\).
In the KPZ phase, \(q_{\rm ss}>0\) but \(q_{\rm st}=0\): 
 samples agree with one
another, but not with the true configuration.
Successful inference has
\(q_{\rm st}>0\), which also implies \(q_{\rm ss}>0\).

The phase boundary separating the two unpinned phases is the bulk roughening transition of the 
height field, or equivalently the 
weak-to-strong disorder transition for the
ordinary directed polymer in a random environment. Let $g_*$ denote its critical bulk-disorder strength. Since the teacher 
has no effect on the 
extensive part of the bulk free energy in either of the unpinned phases, 
the location of this transition depends only on $g_{ss}$ and not on $g_{st}$. The EW-unpinned--to--KPZ-unpinned phase boundary (blue line in the Figure) is therefore  a segment of the vertical line
\[
    g_{ss}=g_*.
\]

So far we have established the  existence of the three stable phases, and the fact that the EW--KPZ boundary is vertical (without assuming the $\epsilon$-expansion for the nontrivial fixed points).
Some basic properties of the two other  phase boundaries can also be identified.
At the KPZ-unpinned--to--pinned transition (purple line), the long  excursions of the candidate polymer away from the pin 
obey the usual KPZ dynamical scaling, with exponent $z_{\rm KPZ}<2$. 
The diffusive motion of the teacher is therefore asymptotically irrelevant at this transition.
Therefore we expect the transition  to lie in the universality class of a DPRE depinning from a static columnar defect~\cite{LassigKinzelbach1997}.

The EW--pinned transition is different. Since bulk disorder is irrelevant throughout the EW phase, the RG flow is toward the $g_{ss}=0$ pinning transition, described by the ``random-walk pinning model'' of a diffusive polymer binding to a (quenched) diffusive random pin~\cite{berger:hal-01426326,10.1214/09-AIHP319,AIHPB_2011__47_1_259_0,Giacomin2007-ns}. 
The critical properties of this boundary are more subtle, and will be discussed in Sec.~\ref{sec:infinite-order-transitions}.

We next consider where these three phase boundaries meet. Within the $\epsilon$-expansion, the multicritical fixed point lies at $(g_{ss},g_{st})=(\epsilon,\epsilon)$, and hence on the Bayes-optimal line. This consequence is not merely an artifact of the $\epsilon$ expansion.
Bayes optimality enhances the $S_{n-1}$ permutation symmetry among the student replicas to an exact $S_n$ symmetry that includes the teacher. Thus the teacher is statistically equivalent to another replica.
Heuristically, this suggests that, on the Bayes optimal line, the pinning and the bulk disorder both become critical at the same time.

The result that the multicritical point lies on the Bayes-optimal line can be established more carefully.
Let $(g_{\rm BO}, g_{\rm BO})$ denote the depinning threshold along the Bayes optimal line. We wish to show that this is the multicritical point, or equivalently that  ${g_{\rm BO} = g_*}$.

First, the inequality $g_{\rm BO}\le g^*$ can be established using a Nishimori-like free-energy identity that holds on the Bayes-optimal line.
As a simplification which does not change the basic point, we will give the identity for the continuum model, in which higher cumulants of the measurement distribution are dropped.
Let $\hat F(g_{ss},g_{st})$ denote the quenched free energy 
of the candidate polymer, averaged over measurement records and
with a trivial analytic contribution removed (see App.~\ref{app:BO_free_energy}). 
Bayes optimality gives the exact relation
\begin{equation}
    -\partial_{g_{st}} \hat F
    =
    2\,\partial_{g_{ss}}\hat F
    \qquad (\text{when  } g_{st}=g_{ss}),
\label{eq:BO_free_energy_identity}
\end{equation}
derived in Appendix~\ref{app:BO_free_energy}.

Suppose, contrary to the phase diagram in Fig.~\ref{fig:3D_main}(a), that ${g_{\rm BO}>g_*}$. The segment ${g_*<g<g_{\rm BO}}$ of the Bayes-optimal line would then lie in the KPZ-unpinned phase. Since the candidate polymer is unpinned from the teacher, the extensive part of its excess free energy is purely the bulk contribution,
\begin{equation}
    \hat F(g_{ss},g_{st})=t\,\mathfrak f\,(g_{ss})+o(t),
\end{equation}
where $\mathfrak f$ is the bulk DPRE free-energy density. Hence $\partial_{g_{st}} \hat F=o(t)$. Eq.~\ref{eq:BO_free_energy_identity} then requires $\partial_{g_{ss}}\hat F=o(t)$, and therefore $\mathfrak f\,'(g)=0$ throughout $g_*<g<g_{\rm BO}$. This is impossible: $\mathfrak f\,(g_*)=0$, whereas $\mathfrak f\,(g)>0$ throughout the KPZ phase. We therefore have
\[
    g_{\rm BO}\leq g_*.
\]

Ruling out the opposite possibility, $g_{\rm BO}<g_*$, requires stronger information about the ordinary $3+1$D DPRE. In Sec.~\ref{sec:infinite-order-transitions} we use recent rigorous results on the weak-disorder phase and its critical point, together with a relation between Bayes-optimal tracking and the so-called \emph{size-biased} ensemble, to show that the candidate polymer cannot acquire an extensive binding free-energy gain before bulk roughening. This gives the complementary inequality $g_{\rm BO}\geq g_*$ (Eq.~\eqref{eq:gBOlower}). Combining the two results gives $g_{\rm BO}=g_*$, so the EW-unpinned, KPZ-unpinned, and pinned phases meet at
\[
    (g_{ss},g_{st})=(g_*,g_*),
\]
on the Bayes-optimal line.

We study the three depinning transitions numerically in Sec.~\ref{sec:critproperties3D}, while Sec.~\ref{sec:infinite-order-transitions} discusses the unusual critical properties of the bulk and Bayes-optimal critical points.

{
\subsection{Misspecification curve in 3D}

A natural path through this phase diagram is model
misspecification: fix the true observation process (i.e.\ $\Delta$),
and vary only the assumed measurement strength, $\Delta'$.
Since $g_{ss}\propto(\Delta')^2$ while
$g_{st}\propto\Delta\Delta'$, this traces out the curve
\begin{equation}
    g_{st}=\frac{\Delta}{2}\sqrt{g_{ss}}.
\end{equation}
The black dotted line in Fig.~3(a) shows the particular misspecification curve that passes through the multicritical point.

A basic fact is that  introducing misspecification (at a fixed $\Delta$) cannot turn unsuccessful tracking into successful tracking.
(Formally,  Bayes-optimal tracking maximizes the expected log score of the true trajectory~\cite{Gneiting2007,Good1952}; see Supplemental Material.)
Therefore, no misspecification curve that crosses the Bayes-optimal line in the unpinned phase can enter the pinned phase.
In particular, the boundary between the KPZ-unpinned and pinned phase 
cannot lie below the misspecification curve through the multicritical point. This is why the purple line in Fig.~\ref{fig:3D_main}(a) lies above the dotted black curve.

The inset to Fig.~\ref{fig:3D_main}(d) shows another misspecification path through the phase diagram (dashed orange line), but now for a larger value of $\Delta$, so that at Bayes optimality tracking is successful.
Interestingly, this path enters all three phases.
 Starting from Bayes optimality \({\kappa\equiv \Delta'/\Delta=1}\) in the pinned phase, 
 the two directions of misspecification lead to the two failure modes. 
 Sufficient underconfidence  underweights the measurements and drives the posterior into the EW-unpinned phase (small $\kappa$) while  sufficient overconfidence  overweights them giving the KPZ-unpinned phase (large $\kappa$). The resulting nonmonotonic 
 behavior of $\ell_{\mathrm{rms}}$ using retrospective tracking is shown in Fig.~\ref{fig:3D_main}(d).

}

\begin{figure} \centering 
\includegraphics[width=1.0\linewidth]{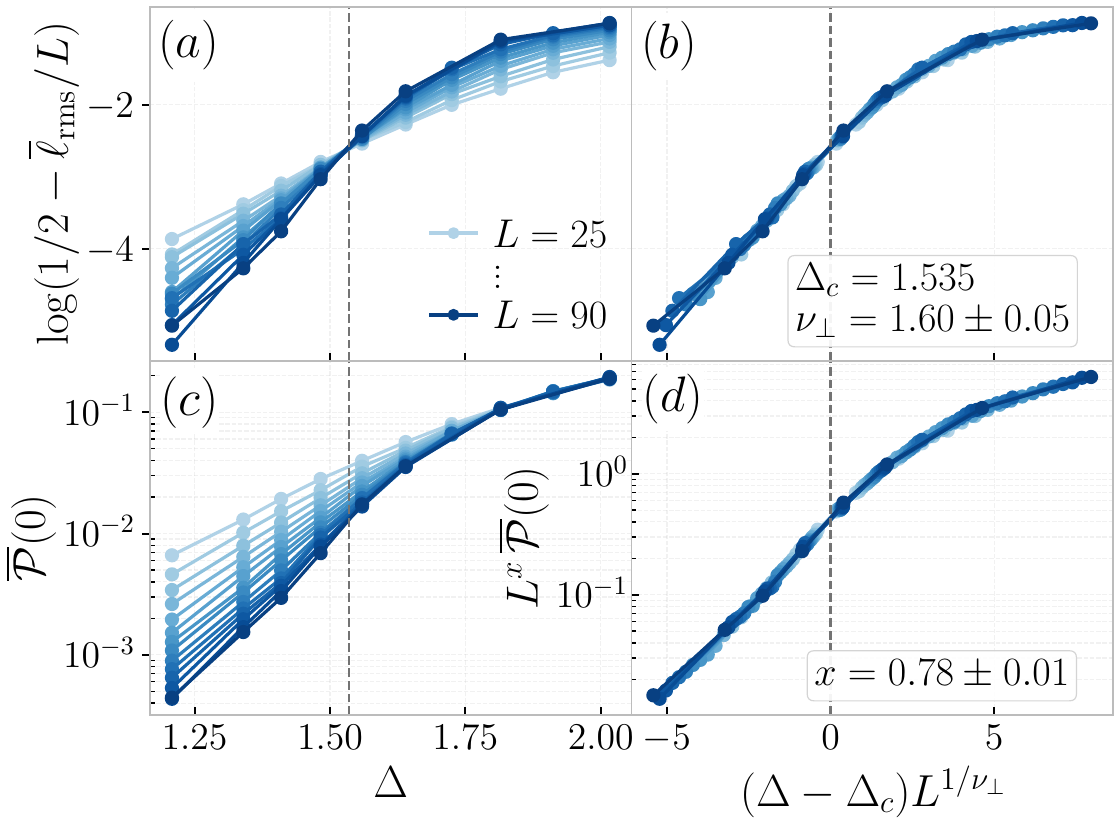} 
\caption{
\textbf{KPZ--pinned transition in \(d=3\).} Finite-size scaling for the KPZ--pinned diagnostic scan of Fig.~\ref{fig:3D_main}(c).
{\bf (a)} the quantity
\(\log(1/2-\overline{\ell_{\rm rms}}/L)\), whose curves exhibit a clear common crossing at \(\Delta_c\simeq1.535\).
{\bf (b)} collapse of the same observable as a function of \((\Delta-\Delta_c)L^{1/\nu_\perp}\), giving \(\nu_\perp=1.60(5)\).
{\bf (c)} the contact fraction \(\overline{\mathcal P}(0)\).
{\bf (d)} contact-fraction collapse using the same \(\Delta_c\) and \(\nu_\perp\), with \(\overline{P(0)}=L^{-x}\mathcal F[ (\Delta-\Delta_c)L^{1/\nu_\perp}]\), giving \(x=0.78(1)\). The scan varies the true measurement contrast \(\Delta\) at fixed assumed model, with the same parameters and numerical protocol as the KPZ--pinned scan in Fig.~\ref{fig:3D_main}(c); further details are given in Appendix~\ref{app:numerics}.
}
\label{fig:3D_KPZ}
\end{figure}

{
\section{Depinning transitions in $d=3$: numerical results}
\label{sec:critproperties3D}

We probe the three depinning transitions---the EW--to--pinned, KPZ--to--pinned, and Bayes-optimal depinning transitions---using the diagnostic scans indicated in Fig.~\ref{fig:3D_main}~(a). The Bayes-optimal scan varies the true and assumed models together and crosses the multicritical point $(g_*,g_*)$. The other two scans vary the true observation strength at fixed assumed model, giving vertical cuts through the $(g_{ss},g_{st})$ plane: one crossing the EW--pinned boundary at $g_{ss}<g_*$, and one crossing the KPZ--pinned boundary at $g_{ss}>g_*$. We do not study the transition between the two unpinned phases here, since this is the standard bulk roughening transition of the DPRE. Figure~\ref{fig:3D_main}(c) shows finite-size scans of $\ell_{\rm rms}/L$ along these three paths.

We first consider the KPZ--pinned transition, whose critical behavior is more conventional. 
As argued in Sec.~\ref{sec:tracking3D}, the diffusive motion of the teacher is asymptotically irrelevant at the KPZ--pinned transition, so we expect the same universality class as for depinning of a $3+1$D directed polymer from a \textit{static} columnar defect~\cite{LassigKinzelbach1997}.
The critical exponents of this $3+1$D transition do not appear to have
been determined previously.

Figure~\ref{fig:3D_KPZ} shows the corresponding finite-size analysis.
In the unpinned phase, $\overline{\ell}_{\rm rms}/L\to1/2$, which compresses the raw curves near their limiting value. The quantity $\log(1/2-\overline{\ell}_{\rm rms}/L)$ magnifies differences between the different finite size curves, and exhibits a clear common crossing. A conventional scaling collapse gives
\begin{equation}
    \nu_{\perp,\mathrm{KPZ}}=1.60(5).
\end{equation}
An independent estimate from the finite-size dependence of the crossing slopes gives the consistent value $\nu_{\perp,\mathrm{KPZ}}=1.57(5)$; see Appendix~\ref{app:numerics}.  The errors we quote are purely statistical errors from the fitting protocol: finite size errors may be larger (see
Appendix~\ref{app:numerics} for the details on the fitting protocol).

For the contact fraction we use the conventional scaling form $\overline{\mathcal P(0)}=L^{-x} \mathcal F\!\left[(\Delta-\Delta_c)L^{1/\nu_\perp}\right]$, which gives
\begin{equation}
    x_{\mathrm{KPZ}}=0.78(1).
\end{equation}
The dynamical exponent does not enter these static scaling collapses. As in $d=2$, it is set by the long unpinned bulk excursions of the polymer, giving $z=z_{\rm KPZ}\simeq1.69$. Together with this value, the results for $\nu_{\perp,\mathrm{KPZ}}$ and $x_{\mathrm{KPZ}}$ provide an independent test of the scaling relation derived in the following section.

For the EW--pinned and Bayes-optimal transitions, $\overline{\ell}_{\rm rms}/L$ likewise shows finite-size crossings, providing estimates of the corresponding transition points. However, the theoretical discussion in Sec.~\ref{sec:infinite-order-transitions} strongly suggests that the asymptotic critical behavior of these transitions is not described by conventional finite-size scaling. In view of this, it is perhaps surprising that conventional scaling collapses nevertheless give a good description of the data over the accessible range of system sizes. For the EW--pinned transition we obtain
\[
    \nu_{\perp,{\rm EW}}^{\rm eff}=1.44(2),
    \qquad
    x_{\rm EW}^{\rm eff}=1.20(1),
\]
while along the Bayes-optimal scan we find
\[
    \nu_{\perp,{\rm BO}}^{\rm eff}=1.78(8),
    \qquad
    x_{\rm BO}^{\rm eff}=1.20(2).
\]
We label these as effective exponents: although the collapses are reasonably good, the available system sizes span well under a decade, and we do not assume that these scaling forms describe the asymptotic regime. Further details and plots are given in Appendix~\ref{app:numerics}.

{
\section{Scaling relation for pinning transitions}
\label{sec:contactfrac}

We now argue that the contact-fraction exponent $x$ is determined by $\nu_\perp$ together with the bulk properties of the height field. Consider the case where the bulk is KPZ rough (this applies both to the pinned-to-KPZ-unpinned tracking transitions above, and to the unbinding of a DPRE from a columnar defect \cite{LKL,tang1993directed,balents1994disorder,hwa1995disorder, LassigKinzelbach1997}). Fixing both the pin configuration and the bulk disorder $\eta$, let the average time-integrated contact between the candidate polymer and the pin be
\begin{equation}
    \langle N_t\rangle
    =
    \int_0^t d\tau\,\mathcal P_\tau(0),
\end{equation}
written here in the pin-centered frame. In the polymer formulation, a perturbation $\varepsilon$ of the defect attraction away from its critical value is conjugate to $N_t$: if $Z_\varepsilon(t)$ is the corresponding polymer partition sum and $F(\varepsilon,t)=-\overline{\log Z_\varepsilon(t)}$ the quenched free energy, then
\begin{equation}
    \overline{\langle N_t\rangle_\varepsilon}
    =
    -\partial_\varepsilon F(\varepsilon,t).
\end{equation}

At a conventional finite-order depinning transition, the singular part of the free energy takes the scaling form
\begin{equation}
    F_{\rm sing}(t,L,\varepsilon)
    \sim
    t^{\beta'}
    \mathcal F\!\left(
        \varepsilon L^{1/\nu_\perp},
        t/L^z
    \right).
\end{equation}
For a KPZ-rough bulk, $\beta'=\beta_{\rm KPZ}$, the exponent setting the scale of bulk free-energy fluctuations. This is required, for example, for consistency with KPZ scaling in the unbound phase.\footnote{Consider fixing a small negative $\varepsilon$ in the unbound phase and taking $t,L\to\infty$ with $t=\mathcal O(L^z)$. In this limit $\mathcal F$ becomes independent of $\varepsilon$, because the expected number of visits of the polymer to the pin is negligible. The free-energy fluctuations must then reduce to the bulk KPZ scaling form $F_{\rm sing}=t^{\beta_{\rm KPZ}} \widetilde{\mathcal F}(t/L^z)$. This requires both that $\mathcal{F}(u, v)$ becomes independent of $u$ when $u$ is large and negative, and that $\beta' = \beta_\text{KPZ}$.} Taking the large-$t$ limit, the scaling form together with extensivity of the free energy gives ${F_{\rm sing}\sim t\,L^{z_{\rm KPZ}(\beta_{\rm KPZ}-1)}\mathcal G(\varepsilon L^{1/\nu_\perp})}$.
Differentiating with respect to $\varepsilon$, and using $\overline{\mathcal P(0)} \sim\overline{\langle N_t\rangle}/t\sim L^{-x}$ at criticality, gives
\begin{equation}
    x+\frac{1}{\nu_\perp}
    =
    z_{\rm KPZ}(1-\beta_{\rm KPZ}).
    \label{eq:contact_scaling_relation}
\end{equation}

In \(d=2\), the large-scale zero-temperature simulations of the static columnar-defect problem in Ref.~\cite{LKL} find exponents \(x\simeq0.66\) and \(\nu_\perp=1.80\). Using the standard \(2+1\)D KPZ estimates \(z\simeq1.61\) and \(\beta\simeq0.24\) \cite{PhysRevA.44.2345,pagnani2015numerical,PhysRevE.84.061150, 2015JSP...160..794H,PhysRevLett.109.170602,PhysRevLett.58.2087}, Eq.~\eqref{eq:contact_scaling_relation} predicts $x_{\rm pred}=z(1-\beta)-\nu_\perp^{-1}\simeq0.668$, in excellent agreement with Ref.~\cite{LKL}. This is also consistent with our $d=2$ diffusing-target estimate of the contact exponent.

The same relation gives an independent check on the KPZ--pinned transition in \(d=3\). Using \(\beta\simeq0.189\), \(z\simeq1.69\)~\cite{PhysRevLett.58.2087,PhysRevE.93.052131,PhysRevE.81.031112, 2000JPhA...33.8181M,PhysRevE.106.L062103}, together with our estimate \(\nu_{\perp,{\rm KPZ}}=1.60(5)\), gives $x_{\rm pred}=0.75(2)$, in good agreement with the directly measured $x_{\rm KPZ}=0.78(1)$. Thus Eq.~\eqref{eq:contact_scaling_relation} quantitatively relates the critical contact probability to the geometry and free-energy fluctuations of the unpinned bulk polymer.

The derivation above assumes conventional finite-size scaling with a finite $\nu_\perp$. As discussed in Sec.~\ref{sec:infinite-order-transitions}, there are strong theoretical reasons to expect different asymptotic behavior for the EW--pinned and Bayes-optimal transitions. Nevertheless, over the accessible range of system sizes both transitions are well described by conventional finite-size scaling. It is therefore useful to ask whether the corresponding effective exponents satisfy the scaling relation above.

For the EW--pinned transition, if the apparent conventional scaling were asymptotic, the same derivation would go through with $z=2$ and $\beta'=0$, since the unpinned bulk is diffusive and its free-energy fluctuations remain $O(1)$ at long times. The resulting relation would be $x+\nu_\perp^{-1}=2$. Using $\nu_{\perp,{\rm EW}}^{\rm eff}=1.44(2)$ gives $x_{\rm pred}^{\rm eff}=1.31(1)$, compared with the measured $x_{\rm EW}^{\rm eff}=1.20(1)$.

At the Bayes-optimal multicritical point the same comparison is more subtle, because depinning coincides with the bulk roughening transition. If the depinning scaling were conventional,  
the same argument would again apply with $z=2$ and $\beta'=0$, since $\beta'=0$ for a finite-temperature fixed point \cite{DotyKosterlitz1992}. Using $\nu_{\perp,{\rm BO}}^{\rm eff}=1.78(8)$ then gives $x_{\rm pred}^{\rm eff}=1.44(3)$, compared with the measured $x_{\rm BO}^{\rm eff}=1.20(2)$. Thus the scaling relation, which works quantitatively for the KPZ--pinned transitions, shows a modest discrepancy for the EW--pinned transition and a more substantial one at Bayes optimality. 
}

Assuming the latter transitions are indeed infinite order transitions (Sec.~\ref{sec:BOtransition}, Sec.~\ref{sec:ewpinnedasymptotics}), then we conjecture that $x=2$ (as obtained from the scaling relation by formally taking $\nu_\perp =\infty$ and $z=2$).

\section{Generalizing to a misspecified diffusion constant}
\label{sec:misspecificationofD}

Above, the diffusion constant $D'$ used by the inferrer was equal to $D$. More precisely, we assumed that the inferrer knew the full hopping kernel $K(x|y)$; however for the present discussion only the diffusion constant $D'$ associated with the inferrer's kernel is important.

The general topology of the phase diagrams in the $(g_{ss}, g_{st})$ plane 
remains the same if we set  ${D'\neq D}$, though in this case there is no longer a Bayes-optimal line that plays a privileged role. 
In fact, the weak-coupling RG equations retain precisely the  form  in Eq.~\ref{eq:generic_beta} if we define ${g_{ss} =\frac{K_d}{2D'}\Lambda^{d-2}\mathsf g_{ss}}$ and 
 ${g_{st} =\frac{K_d}{D+D'}\Lambda^{d-2}\mathsf g_{st}}$, so that the weak-coupling analysis  also remains similar (see Supplemental Material). 
 For example, we again have an exponentially growing lengthscale within the 2D pinned phase at weak coupling.

The universality class of the transition between the \textit{KPZ  phase} and the pinned phase is insensitive to the ratio  $D'/D$, both in 2+1D and in 3+1D, 
for a reason already discussed: the motion of the teacher is RG-irrelevant at this type of transition.

For the other transitions out of the pinned phase in 3+1D, 
where the dynamical exponent is ${z=2}$,  the parameter $D'/D$ is formally an exactly marginal variable, because it is invariant under a $z=2$ rescaling.

\section{Repairing nonoptimal Bayesian inference}
\label{sec:repairingnonoptimal}

At first one might think that \textit{optimal} inference was only relevant to the case where the inferrer has prior knowledge of $D$ and $\Delta$. But the Bayes-optimal phase diagram has a broader relevance, because instead of holding  the assumed parameters $D'$ and $\Delta'$ fixed, the inferrer may instead be able to estimate them from the measurement data $\{\eta\}$ (cf. the spin model setting \cite{NahumBayesianCriticalPoints}).
We may argue that if  the inferrer is allowed to scan over values $(D', \Delta')$, and has access to data over long enough timescales, then they can succeed in inference if and only if the \textit{true} parameters are such that Bayes-optimal inference would succeed.

The point is that the inferrer can detect from the behavior of the posterior whether or not they are in the successful-tracking phase (see the Supplemental Material for more details). This allows them to scan parameters until they lock on to the true trajectory. 

The phase diagram for \textit{non}-optimal inference nevertheless remains important for practical inference tasks.  First, the size of the pinned phase, within this phase diagram,
determines the ease of the above scanning task. 
Second, scanning over/inferring the assumed parameters may be impractical in many more realistic settings: for example because the assumed model has too many parameters, or because computing power is limited, or because the amount of available data is too small. Similar considerations are relevant to other nonoptimal inference tasks.

\section{Phase transitions for directed polymers in $d+1$ dimensions}
\label{sec:infinite-order-transitions}

In this Section we discuss the critical properties of various phase transitions out of the Edwards-Wilkinson (EW) phase, in 3+1D and above.
These include (I) the transition into the KPZ phase, which we refer to  as the ``bulk'' roughening transition, since no pin is involved; (II) 
the EW-to-pinned transition; 
and (III) the transition on the Bayes-optimal line.
These transitions are relevant well beyond the present inference problem. For example, the bulk roughening transition appears in surface growth models, and we expect the EW-to-pinned transition to share universal properties with the random walk pinning model.

The discussion here is partly an initial response to recent rigorous results~\cite{junk2024strong,Junk2025coincidence,lacoin2025localization} showing that the bulk transition (I) is infinite-order (contrary to earlier expectations in the physics literature~\cite{derrida1990thermal})  and to  results suggesting or demonstrating infinite-order transitions in some other pinning problems~\cite{tang2001rare,derrida2014depinning,chen2021derrida}. 
They are also a response to the fact that the conventional $\epsilon$ expansion fails for these transitions.
We will suggest that transitions (I)--(III) are governed by an unusual RG structure, in which a single RG fixed point controls both the stable EW phase and several distinct transitions out of it, with the different cases distinguished by the flow lines along which this fixed point is approached.

For the tracking problem, the bulk results have an additional consequence. Combined with exact identities following from Bayes optimality, they allow us to show that the Bayes-optimal depinning transition occurs precisely at the bulk roughening threshold, as stated in Sec.~\ref{sec:tracking3D}. The same connection also constrains the critical properties of this multicritical point. 
For the purposes of the inference problem, the conclusion is that the transitions of type (II) and (III) are expected to be infinite-order transitions, with the candidate polymer becoming asymptotically a free random walk at the transition point.
To begin with, however, we must discuss the status of the bulk transition (I).

\subsection{Bulk roughening transition}
\label{sec:bulkroughening}

Let us start with the bulk roughening transition. The $\epsilon$ expansion around ${d=2}$ gives the \textit{putatively} exact result ${\nu=1/(d-2)}$ for this transition, since the beta function truncates at quadratic order.  
However, this value is not in good agreement with numerics in 3+1D 
\cite{derrida1990thermal,kim1991finite,MonthusGarel}
(as well as being incompatible with a rigorous bound  when $d>4$ \cite{cardy1999field}; the stability  of the $\epsilon$ expansion to higher-order interactions was also questioned~\cite{nakayama2021efimov}).

Recent rigorous work by Junk and Lacoin shows that the bulk critical point is very different from the $\epsilon$ expansion picture \cite{junk2024strong,junk2024tail,Junk2025coincidence,lacoin2025localization}. The  correlation length exponent is in fact infinite, ${\nu=\infty}$, and right at the critical point, the polymer is (asymptotically) a simple Brownian path. That is, quenched randomness is asymptotically irrelevant at  critical point (I).

Since this picture is different from that previously expected by physicists it is interesting to ask whether it can be rationalized using the RG. In Sec.~\ref{sec:rgcomments} below we will make some brief points about this (which we will discuss more fully elsewhere).  We note that an infinite-order transition can be  obtained from a real-space RG for the tail of the disorder distribution. We also point out that a slight extension of the conventional perturbative RG in fact hints at the infinite order transition.

First, however, let us summarize some basic findings of Refs.~\cite{junk2024strong,junk2024tail,Junk2025coincidence,lacoin2025localization}---which focus on the statistics of the DPRE partition function---in crude terms. (We refer to the papers for precise statements.)
 
Let $\mathcal{Z}_{t;\eta}$ be the ``point-to-volume''\footnote{I.e. with the initial point fixed and the final point free.} partition function for a DPRE of length $t$, in a spatiotemporal disorder realization $\eta$. 
It is convenient to absorb a trivial factor into $\mathcal{Z}_{t;\eta}$ so that $\overline{\mathcal{Z}_{t;\eta}}=1$, where the average is over~$\eta$.

With this normalization, the free energy  density $\mathfrak{f}$
(with ${\mathfrak{f} \equiv - \lim_{t\to\infty} t^{-1}\,\overline{ \ln \mathcal{Z}_{t;\eta}}}$)
\textit{vanishes} in the  EW phase and at the roughening transition, i.e. for $g_{ss}\leq g_{ss}^*$.
In other words, ${\mathfrak{f}}$ is equal to its annealed value in this regime. 
By contrast $\mathfrak{f}$ is positive and varies with $g_{ss}$ within the KPZ phase $g_{ss}>g_{ss}^*$.\footnote{Recall that 
in our microscopic model the inverse temperature for the  polymer is $\Delta' = 2\sqrt{g_{ss}}$  (Eq.~\ref{eq:worldinedisorder}).}

Remarkably, the statistics of the partition function right at the critical point $g_{ss}^*$ are well-characterized~\cite{junk2024strong,junk2024tail}. 
In the limit of large $t$,
$\mathcal{Z}_{t;\eta}$ converges
to a nonzero random variable $\mathcal{Z}_\infty$ with a well-defined probability distribution.\footnote{This is referred to in the mathematical literature as the ``weak disorder'' property. It holds in the EW phase and at the critical point. By contrast in the KPZ phase ``strong disorder'' holds which means that ${\mathcal{Z}_\infty=0}$ with probability 1. (Physically, the latter holds because $\mathfrak{f}>0$ in the KPZ phase, meaning that the partition function is typically exponentially small at large $t$.)} 
The tail of this probability distribution at large $\mathcal{Z}_\infty$ scales as\footnote{The precise statement in Ref.~\cite{junk2024tail} involves upper and lower bounds of order $\mathcal{Z}^{-(1+2/d)}$.}
\begin{equation}
\Pr(\mathcal{Z}_\infty>\mathcal{Z}) \propto {\mathcal{Z}^{-(1+2/d)}}.\label{eq:Ztail}
\end{equation}

Diffusive behavior at sufficiently weak disorder was established already in Ref.~\cite{Imbrie1988}, and later proved to hold throughout the weak-disorder phase $g_{ss}<g_{ss}^*$~\cite{comets2006directed} (as expected from the RG). Remarkably, this Brownian behavior persists also at the critical point $g_{ss}=g^*_{ss}$~\cite{junk2024strong}. That is, after rescaling the time coordinate by $t$ and the space coordinate by $\sqrt{t}$, the polymer trajectory converges at large $t$ to a Brownian motion with rescaled time coordinate in $[0,1]$.

Finally, the statistics of the partition function have been used to show that the correlation length exponent at the transition diverges. For a ``conventional'' finite-order transition (governed by a finite-temperature RG fixed point~\cite{DotyKosterlitz1992} at nonzero disorder strength) the scaling form for the free energy (Sec.~\ref{sec:contactfrac}) would yield ${\mathfrak{f} \sim (g_{ss}-g_{ss}^*)^{z\nu_\perp}}$ just inside the KPZ phase. 
Instead Ref.~\cite{lacoin2025localization} shows that $\mathfrak{f}$ vanishes faster than any power law as the transition is approached.

\subsubsection{Comments on RG: rare regions and a modified $\epsilon$~expansion}
\label{sec:rgcomments}

A transition with ${\nu = \infty}$ can be obtained from  an approximate real-space RG if we assume that it is sufficient to focus on the \textit{tail} of the probability distribution $p(u)$ for the local Boltzmann weight ${u = e^{-V}}$ on a bond.  Here we make some brief comments about an approach which we will describe in full elsewhere~\cite{NahumInPrep}.
This is inspired by related RG approaches that have been proposed for the disordered renewal model (pinning to a static pin)  in Refs.~\cite{tang2001rare,derrida2014depinning,monthus2017strong}.\footnote{Infinite-order transitions have also been discussed in the context of many-body localization~\cite{PhysRevLett.122.040601,PhysRevB.99.094205,PhysRevB.102.125134}.}

First, simple rare-region arguments show that, within the EW phase, $p(u)$ develops a power law tail, at large $u$, under RG. A simple approximate RG then considers the amplitude $A$ and exponent $\alpha$ for this tail, under the assumption that the transition is driven by the tail \cite{NahumInPrep}. 
We find flows in the $(\alpha,A)$ plane that are reminiscent of Kosterlitz-Thouless flows, yielding an infinite order transition at a critical value ${\alpha=1+2/d}$ that is consistent with the results of Refs.~\cite{junk2024strong,junk2024tail}. 

An unusual feature of this RG is that, formally, both the  EW phase and the transition out of it are governed by the same RG fixed point. The critical point, and points within the phase, 
differ only through the \textit{flow line} along which this fixed point is approached.\footnote{The flow lines end on the line ${A=0}$, with different values of $\alpha$. At first sight this seems similar to the  Kosterlitz-Thouless flows for the 2D XY model, where the stiffness varies continuously in the quasi-long-range ordered phase. But  in the present case (in contrast to the XY setting), the entire line ${A=0}$ corresponds to a single RG fixed point.
This situation is interesting because it shows that the RG mechanism for a continuous phase transition can differ both from the standard one where a continuous phase transition has its own ``critical'' fixed point, and the Kosterlitz-Thouless mechanism, which involves an RG fixed line.}
In both the EW phase and at the roughening transition, the polymer is asymptotically Brownian, i.e. 
at large scales it has the statistics of an ordinary directed random walk, for typical disorder realizations, in agreement with the rigorous results.

Finally, we argue that an extension of the ``conventional" field theoretic RG approach to multiple numbers of replicas also hints at the divergence of $\nu$.

The replica method studies the moments $\overline{\mathcal{Z}^k}$~\cite{EvansDerrida1992,Derrida2009,KARDAR1987582}, where $\mathcal{Z}$ is the partition function for a DPRE and the average is over bulk disorder. Defining ${F_k = - k^{-1} \ln \overline{\mathcal{Z}^k}}$,  the limit ${k\to 0}$ gives the quenched free energy.
As a result of averaging,  $\overline{\mathcal{Z}^k}$ (for ${k=2,3,4,\ldots}$) is a partition function for $k$ random walks (or $k$ quantum particles) with pairwise attractions. 

The standard epsilon expansion results are certainly valid for $\overline{\mathcal{Z}^2}$, i.e. for the binding transition of a pair of particles.\footnote{Or equivalently (by separating out the relative coordinate) the binding of a single walker to a static pin.}
However, we may argue that the phase transition of $\overline{\mathcal{Z}^k}$
for $k>2$ is \textit{not} governed by an RG fixed point at nonzero pairwise coupling.\footnote{In Ref.~\cite{nakayama2021efimov}, it was  suggested that the three-replica interaction was relevant for the physics of the roughening transition (above a critical dimensionality). The insufficiency  of considering only two-replica interactions was an important insight, although we do not agree with the specific picture for the transition in Ref.~\cite{nakayama2021efimov}.}   Instead, we argue that, at the critical point for $\overline{\mathcal{Z}^k}$, the {pairwise} coupling flows asymptotically to zero, and  the critical fixed point is one where \textit{only} the  $k$-replica interaction $g_k$  
(which is generated under RG)
is present. See App.~\ref{app:kreplica} for details.
Then, a similar $\epsilon$ expansion, but around the modified critical dimensionality  $d_k = 2/(k-1)$, gives 
\begin{equation}
{\nu_k = \left((k-1)d-2\right)^{-1}}.\label{eq:nucont}
\end{equation}
This exponent would appear, for example, in the scaling form for the $k$-replica free energy ${F_k = - k^{-1} \ln \overline{\mathcal{Z}^k}}$.

Let us assume that $\nu_k$ can be continued from the discrete values ${k\in \{2, 3,4,\ldots\}}$ to some range of $k{\in \mathbb{R}}$.
Note that $\nu_k$ in Eq.~\ref{eq:nucont} diverges as ${k\to 1+2/d}$ from above: therefore the natural conjecture is that (\ref{eq:nucont}) holds for ${k\geq 1+2/d}$, with the transition being infinite order
($\nu_k=\infty$) for smaller $k$.
This would imply that the transition was infinite order for the DPRE (in agreement with \cite{lacoin2025localization}) and also for the Bayes-optimal transition, since these correspond to $k\to 0$ and $k\to 1$ limits of $F_k$ respectively.\footnote{Going further, we suspect that an  analytic continuation of the perturbative RG equations allows  the results of the real-space RG mentioned above to be obtained from the conventional perturbative RG. We will return to this later. Note that the critical $k$ above is also the the critical tail exponent of the partition function, Eq.~\ref{eq:Ztail}.}

\subsection{Transition on the Bayes-optimal line}
\label{sec:BOtransition}

The ``naive'' $\epsilon$-expansion would again suggest a conventional RG fixed point at finite, nonzero disorder strength for this transition, with $\nu=1/(d-2)$.\footnote{From Eq.~\ref{eq:generic_beta} on the symmetric line $g_{ss}=g_{st}$.} There are various reasons for expecting this not to hold. First, even if the assumption of a conventional transition is made,  the resulting finite-size scaling analysis of Sec.~\ref{sec:critproperties3D} gives the effective estimate $\nu_{\perp,\mathrm{BO}}^{\mathrm{eff}}=1.78(8)$, substantially different from the $\epsilon$-expansion prediction $\nu=1$; see also Appendix~\ref{app:numerics}. 
Second, the RG argument above, based on a hierarchy of $\epsilon$ expansions for different replica numbers,  suggests a divergence of the correlation length exponent  upon analytic continuation to the Bayes-optimal replica limit, contrary to the picture of a conventional fixed point at nonzero disorder strength.

More direct information follows from an exact relation between Bayes-optimal tracking and the ordinary DPRE. Interestingly, this invokes the ``size-biased ensemble'' that appears in the mathematical literature  \cite{Birkner2004,zygouras2024directed}.

Let $\mathbb P_0$ denote the probability measure on spacetime environments $\eta=\{\eta_\tau(x)\}$ in which each $\eta_\tau(x)\in\{0,1\}$ is drawn independently from the background distribution with parameter $p_0$. Let $\mathbb P$, without a subscript, instead denote the probability measure on measurement records in the tracking problem, after marginalizing over the true trajectory $X$.

For a \textit{fixed} teacher trajectory $X$, the measurement statistics differ from the background measure $\mathbb P_0$ only at the spacetime points $(X_\tau,\tau)$ visited by the teacher. By the model in Sec.~\ref{sec:model},
\begin{equation}
\frac{\mathbb P(\eta|X)}{\mathbb P_0(\eta)}
=
\left(\frac{1-p_1}{1-p_0}\right)^t
\exp\left[
\Delta\sum_{\tau=1}^t\eta_\tau(X_\tau)
\right],
\end{equation}
where in the second line we used the log-likelihood contrast $\Delta$ defined in Eq.~\eqref{eq:contrasts}.\footnote{In this Section it will be more convenient to use ($\Delta'$, $\Delta$) as our phase diagram coordinates, rather than $(g_{ss}, g_{st})$.} This is the same likelihood-ratio simplification that underlies the forward tracking update rule in Eq.~\eqref{eq:Zupdate}.

Averaging over $X$ with its random-walk prior, the resulting likelihood ratio is proportional to the point-to-volume partition function of the ordinary DPRE at inverse temperature $\Delta$,
\begin{align}
\mathbb E_X \bigg[\frac{\mathbb P(\eta|X)}{\mathbb P_0(\eta)}\bigg] &= \mathcal{Z}_{t; \eta},
\end{align}
where 
\begin{equation}
    \mathcal {Z}_{t; \eta} = c^t \sum_{x_{t},\ldots, x_1}
 e^{\Delta\sum_{\tau=1}^t \eta_\tau(x_\tau)}
\prod_{\tau=1}^{t}
K(x_\tau|x_{\tau-1})
\end{equation}
and ${c \equiv \frac{1-p_1}{1-p_0}}$. 
Moreover, this likelihood ratio has unit average under $\mathbb P_0$:
$\mathbb E_0[\mathbb P(\eta)/\mathbb P_0(\eta)]
=\sum_\eta \mathbb P(\eta)=1$.
Therefore $\mathcal{Z}_{t;\eta}$ is precisely the  DPRE partition function introduced in Sec.~\ref{sec:bulkroughening}, normalized such that ${\mathbb{E}_0 \mathcal{Z}_{t;\eta}=1}$.
We therefore have 
\begin{equation}
\mathbb P(\eta)
=
\mathcal Z_{t;\eta}\, \mathbb P_0(\eta).
\label{eq:size_bias}
\end{equation}

Finally, it remains to connect the DPRE partition function $\mathcal{Z}_{t;\eta}$ above to  the  partition function of the candidate polymer in tracking problem.
Let $Z_{t;\eta}$ denote the point-to-volume candidate partition function obtained by summing $Z_t(x)$ in Eq.~\ref{eq:worldinedisorder} over its final endpoint, with the same normalization convention as above. 

For a fixed measurement record $\eta$, $Z_{t;\eta}$ depends only on $\eta$ and not separately on the hidden teacher trajectory $X$. 
The two partition functions $Z_{t;\eta}$ and $\mathcal{Z}_{t;\eta}$ 
have precisely the same random-walk weights and the same disorder realization $\eta$, but in general differ in their inverse temperatures: $\mathcal Z_{t;\eta}$ is defined with the true contrast $\Delta$, whereas $Z_{t;\eta}$ is defined with the assumed contrast $\Delta'$. However, at Bayes optimality, $\Delta'=\Delta$, and hence $Z_{t;\eta}=\mathcal Z_{t;\eta}$. 
We nevertheless retain the two symbols below when useful, to emphasize whether this same random variable is viewed in the tracking ensemble $\mathbb P$ or in the ordinary DPRE disorder ensemble $\mathbb P_0$.

The identity (\ref{eq:size_bias}) is significant   because it  relates observables at Bayes-optimal tracking to observables for the DPRE by a simple reweighting.
This distribution on the left is referred to in the mathematical DPRE literature as the size-biased measure \cite{Birkner2004,zygouras2024directed}. 
This structure is in fact  a natural consequence of the structure of  optimal Bayesian inference, quite generally. 
In the replica language, the identity (\ref{eq:size_bias}) is associated with the fact that the tracking problem is described by an ${n\to 1}$ replica limit while the DPRE is described by an ${n\to 0}$ replica limit.

We now consider Bayes-optimal tracking at the bulk roughening threshold. Eq.~\ref{eq:size_bias} provides a direct relation between the Bayes-optimal tracking ensemble and the ordinary DPRE, allowing us to transfer information about the critical DPRE to the tracking problem. We use this relation to establish two basic properties.

First, the normalized candidate free energy remains \(O(1)\) at the bulk roughening threshold. This shows that the free energy density vanishes, completing the argument in Sec.~\ref{sec:3dtopology} that the multicritical point lies on the Bayes-optimal line.

Second, despite the size biasing of the disorder ensemble, the critical candidate polymer remains asymptotically a free random walk. This shows that the phase transition is not governed by a conventional finite-disorder fixed point, but instead flows to the trivial free (non-disordered) fixed point. This hints that the transition is infinite order, like that discussed in the previous section.

%We now evaluate Bayes-optimal tracking at the bulk roughening threshold. Equation~\ref{eq:size_bias} allows us to establish two properties. First, the candidate free energy remains \(O(1)\), completing the argument that the Bayes-optimal depinning threshold coincides with bulk roughening. Second, Eq.~\ref{eq:size_bias} allows us to relate critical properties of the ordinary (critical) DPRE directly to the Bayes-optimal critical point.  

%First, the normalized candidate free energy remains $O(1)$ at criticality.
%This shows that the free energy density vanishes, completing the argument in Sec.~\ref{sec:3dtopology} that the multicritical point lies on the Bayes-optimal line.
%Second, despite the size biasing of the disorder ensemble, the critical candidate polymer remains asymptotically a free random walk.
%This shows that the phase transition is not governed by a conventional finite-disorder fixed point, but instead flows to the trivial free (non-disordered) fixed point.
%This hints that the transition is infinite order, like that discussed in the previous section.

We first consider the free energy at the Bayes-optimal critical point. The average of (minus) the candidate's free energy
is
\begin{equation}
\overline{\ln Z_{t;\eta}}
=
\mathbb E_0\!\left[
\mathcal Z_{t;\eta}\ln\mathcal Z_{t;\eta}
\right],
\label{eq:BO_logZ}
\end{equation}
where the overline is the average over tracking instances (equivalently, the average over $\eta$ with the distribution $\mathbb{P}$ above).
Throughout the weak-disorder phase of the ordinary $3+1$D DPRE there
exists ${a>0}$ 
such that~\cite{junk2024tail}
\begin{equation}
\sup_t
\mathbb E_0\mathcal Z_{t;\eta}^{\,1+a}<\infty.
\label{eq:BO_moment_bound}
\end{equation}
For the bounded disorder of our microscopic model, this moment bound continues to hold at the bulk roughening transition~\cite{junk2024strong}.\footnote{The corresponding critical result also holds for unbounded disorder, including Gaussian disorder~\cite{Junk2025coincidence}.} Since $z\ln z\leq z^{1+a}/a$ for $z>1$, while $z|\ln z|\leq e^{-1}$ for $0<z<1$, it follows that, at the Bayes-optimal critical point,
\begin{equation}
\mathbb E_0\!\left[
\mathcal Z_{t;\eta}\ln\mathcal Z_{t;\eta}
\right]
=O(1).
\label{eq:BO_logZ_bound}
\end{equation}
Thus the normalized candidate free energy remains $O(1)$ at criticality: the teacher does not generate an extensive binding free-energy contribution.

Returning to the question of the location of the multicritical point in Sec.~\ref{sec:3dtopology},  and to the phase diagram coordinates $(g_{ss}, g_{st})$ used there, the above shows that the point $(g_*,g_*)$ cannot lie in the pinned phase. This gives the complementary phase-diagram constraint used in Sec.~\ref{sec:tracking3D},
\begin{equation}
g_{\rm BO}\geq g_*.
\label{eq:gBOlower}
\end{equation}

The second critical property is perhaps more striking. Despite the size biasing of the disorder ensemble, the candidate polymer at the Bayes-optimal critical point remains asymptotically a free random walk. To see this, recall that the critical DPRE results imply that, for typical disorder realizations drawn from $\mathbb P_0$, the diffusively rescaled polymer measure converges to Brownian motion~\cite{junk2024strong,comets2006directed}. 
We  show that this property survives the change from $\mathbb P_0$ to the size-biased measure $\mathbb P$.

For any fixed $\epsilon>0$, let $\mathcal B_t(\epsilon)$ denote the set of disorder realizations for which the rescaled polymer measure differs from its Brownian limit by more than $\epsilon$\footnote{Here the distance between probability measures on rescaled trajectories may be taken, for example, to be the bounded-Lipschitz distance~\cite{Dudley_2002}.}. Then
\begin{equation}
\mathbb P_0\!\left(\mathcal B_t(\epsilon)\right)\longrightarrow 0
\qquad (t\to\infty).
\end{equation}
Using Eq.~\eqref{eq:size_bias}, H\"older's inequality, and
Eq.~\eqref{eq:BO_moment_bound},
\begin{align}
\mathbb P(\mathcal B_t)
&=
\mathbb E_0\!\left[
\mathcal Z_{t;\eta}\mathbf 1_{\mathcal B_t}
\right]
\nonumber\\
&\leq
\left(
\mathbb E_0\mathcal Z_{t;\eta}^{\,1+a}
\right)^{1/(1+a)}
\left(\mathbb P_0(\mathcal B_t)\right)^{a/(1+a)}
\longrightarrow0 .
\label{eq:BO_Brownian}
\end{align}
Thus the candidate polymer at the Bayes-optimal multicritical point is asymptotically a free random walk. In RG terms, this shows that the Bayes-optimal critical point flows to the same free fixed point that governs the EW-unpinned phase (and, as discussed above, the bulk roughening critical point). The distinction between these cases lies not in the limiting fixed point itself, but in the flow along which it is approached.

This picture is very different from the finite-coupling fixed point predicted by the naive $\epsilon$ expansion: the Bayes-optimal transition is governed asymptotically by the free-random-walk fixed point, rather than by a nontrivial fixed point at finite disorder and pinning strength. Together with the infinite-order character of the coincident bulk transition and the RG considerations above, this leads us to conjecture that the Bayes-optimal depinning transition is itself infinite order. We therefore expect that the finite $\nu_{\perp,\mathrm{BO}}^{\mathrm{eff}}$ obtained from the numerical scaling collapse in Sec.~\ref{sec:critproperties3D} is a preasymptotic effective exponent.

\subsection{EW-to-pinned transition}
\label{sec:ewpinnedasymptotics}

Finally let us discuss the transition of type (II), 
the transition on the left of the multicritical point (Fig.~\ref{fig:3D_main}(c), left) which separates the Edwards-Wilkinson phase from the pinned phase.
Since bulk disorder is irrelevant at this transition,
we expect this transition to share universal properties with the random walk pinning model,
in which a polymer is attracted to a pin in the shape of a quenched random trajectory (but is not subject to bulk disorder). 
Specifically this is the random-walk pinning model~\cite{berger:hal-01426326,10.1214/09-AIHP319,AIHPB_2011__47_1_259_0,Giacomin2007-ns} in the case where the diffusion constants for the pin and the polymer are equal. 
Therefore the conclusions below are also relevant to this model.

The ``naive'' expectation  would be that this transition (II) and the Bayes-optimal transition
(III) were governed by distinct RG fixed points. 
For example   the standard $\epsilon$ expansion would predict  that  (II) is governed by an RG fixed point at  zero bulk disorder and that (III) is governed by a fixed point at nonzero bulk disorder.

The first of these statements is correct (Sec.~\ref{sec:3dtopology}), but we saw in the previous Section that the second prediction is incorrect: in fact the bulk disorder flows, asymptotically, to zero for both transitions.
This strongly suggests that (II), like (III) is another infinite-order transition, 
at which the polymer is asymptotically Brownian.

This conjecture has the surprising consequence that transitions (I), (II) and (III) are formally all governed by the same---free Brownian---fixed point (see the comment and footnote in Sec.~\ref{sec:rgcomments}).
However, their universal properties 
(for example finite-size scaling forms)
will nevertheless differ as a result of the fact that they are associated with distinct kinds of flow lines into this fixed point.
We will discuss this elsewhere.

\section{Discussion}
\label{sec:discussion}

This work maps out the phase structure for noisy monitoring of a diffusing particle 
\cite{offer2018phase,jin2022kardar,Kim2025} in two and three dimensions, and characterizes the associated depinning transitions.
We summarize some of the key points and then note some extensions and open questions.

The inference problem is a pinning problem for the posterior trajectory ensemble: tracking succeeds when the posterior remains bound to the true trajectory, and fails when it depins. We found that Bayes-optimal tracking is always successful in \(d=2\), although the corresponding localization length may be exponentially large
(consistent with an empirical observation in Ref.~\cite{offer2018phase}), while nonoptimal inference can produce an unsuccessful phase in which the log-posterior has KPZ statistics. In \(d=3\), Bayes-optimal tracking undergoes a phase transition, and the full phase diagram contains two distinct failure phases. Underconfidence (or insufficient information) gives a smooth log-posterior with Edwards--Wilkinson statistics, 
whereas overconfidence produces a KPZ-rough posterior that is localized, but at the wrong position.
Thus model misspecification alone can drive two qualitatively distinct modes of inference failure. They can be
distinguished from each other formally  by the pattern of student-student and student-teacher replica overlaps $q_{ss}$ and $q_{st}$.

The overlap perspective also clarifies the continuously monitored random walker studied in Ref.~\cite{jin2022kardar}. 
In their formulation the latent target trajectory is marginalized out,
so the dynamics is written directly for the posterior and no explicit ``teacher'' or pin appears. 
There is therefore a natural notion of overlap $q_{ss}$ within the posterior ensemble, 
but no explicit teacher--student overlap. 
Reintroducing the latent trajectory makes explicit that, at Bayes optimality, 
the onset of posterior localization,
$q_{ss}>0$,
coincides with localization onto the teacher, $q_{st}>0$. 
We therefore clarify the physical meaning of the strongly rough phase discussed in Ref.~\cite{jin2022kardar} in three dimensions. This is the pinned phase: exponential localization implies \(O(L)\) variation of the log posterior across a system of size \(L\), giving the roughness exponent \(\alpha=1\) noted in Ref.~\cite{jin2022kardar}.

Bayes optimality further organizes the phase diagram.
Enhanced replica symmetry at Bayes optimality suggests that the depinning and bulk-roughening thresholds meet at a multicritical point \((g_*,g_*)\).
More rigorously, this follows from a size-biased representation~\cite{Birkner2004,zygouras2024directed,junk2024tail} of the Bayes-optimal disorder distribution, together with an exact Nishimori free-energy identity.

Bayes optimality also constrains the geometry of the phase boundaries: at fixed true measurement model, the inferrer may vary the model used for inference, but cannot outperform the Bayes-optimal choice. Consequently, if Bayes-optimal tracking fails for a given measurement model, then tracking must also fail for every misspecified inference model.

Turning to the transitions, we obtained a scaling relation for conventional depinning transitions that is in good agreement with numerics for both the \(2+1\)D and \(3+1\)D KPZ--to--pinned transitions. 
The critical properties of the EW--to--pinned and Bayes-optimal depinning transitions remain an interesting open problem: theoretical considerations point towards infinite-order behavior, while the numerical data remain strikingly consistent with conventional finite-size scaling over accessible system sizes.

Let us note a few extensions, starting with some alternative applications of the model.
Although we introduced the longitudinal coordinate as physical time, the
effective statistical-mechanics problem admits other interpretations. 

The coordinate \(t\) may of course instead be regarded as an additional spatial direction,
in which case the task is to reconstruct a directed path or directed polymer
in \(d+1\) spatial dimensions from noisy data
\cite{yuille2002fundamental,Kim2025}. The \(1+1\)-dimensional version has,
for example, been connected to the detection of roads in noisy images
\cite{yuille2002fundamental}. In \(2+1\) dimensions, analogous problems arise
in imaging stretched polymers or extended topological defects in
three-dimensional samples.

A distinct but related polymer problem arises if
\(\{X_\tau\}\) is interpreted as the conformation of a \textit{non-directed}
copolymer in \(d\)-dimensional space, with \(\tau\) being simply the internal coordinate
along the chain. If the subunits are distinguishable and each produces a
separately identifiable noisy signal, one may again ask for the posterior
ensemble of polymer conformations conditioned on the resulting collection
of images. If the conformations of the polymer were ideal (Brownian), we could recover the  effective model studied here.

As a simple generalization of the tracking model studied here, it would be interesting to replace ordinary diffusion by a broader
class of target dynamics with ${|X_t-X_0|\sim t^{1/z}}$ for ${z\neq 2}$. 
Naive dimensional analysis suggests that the local pair-contact interaction should have RG eigenvalue ${y_g=z-d}$ so that $d_c=z$ is the marginal dimension, generalizing \({d_c=2}\) for ordinary diffusion.
Subdiffusive and superdiffusive targets may therefore have qualitatively different tracking phase diagrams even in the same spatial dimension.

Another natural extension would be to active-search problems, in which the observations are not supplied by a fixed distributed sensor array, but depend on the trajectory chosen by a mobile searcher or a team of searchers. Examples include infotaxis, where a searcher adapts its motion to maximize information gain from sparse observations, and proxitaxis, where the search strategy is adapted using proximity information \cite{Vergassola2007,delvecchio2026proxitaxis}. It would be interesting to ask whether analogous tracking phases and transitions arise when inference and the searcher's dynamics are coupled in this way.

\textbf{Acknowledgments}---The authors thank Denis Bernard, Andrea De Luca, Bernard Derrida, Guido Giachetti, Felix Werner, and Kay Joerg Wiese for helpful discussions. EM and AN are supported by the European Union (ERC, STAQQ, 101171399). Views and opinions expressed are however those of the authors only and do not necessarily reflect those of the European Union or the European Research Council Executive Agency. Neither the European Union nor the granting authority can be held responsible for them.

\let\oldaddcontentsline\addcontentsline% Store \addcontentsline
\renewcommand{\addcontentsline}[3]{}% Make \addcontentsline a no-op
\bibliography{Refs}
\let\addcontentsline\oldaddcontentsline% Restore \addcontentsline

\clearpage
\onecolumngrid

\makeatletter
\let\mainaddcontentsline\addcontentsline
\renewcommand{\addcontentsline}[3]{%
  \def\tempa{#1}%
  \def\tempb{toc}%
  \ifx\tempa\tempb
    \mainaddcontentsline{apc}{#2}{#3}%
  \else
    \mainaddcontentsline{#1}{#2}{#3}%
  \fi
}
\makeatother

\makeatletter
\let\set@footnotewidth\set@footnotewidth@one
\onecolumn@grid@setup
\makeatother

\appendix

\makeatletter
\begin{center}
\textbf{\large Supplementary Materials: Bayesian Tracking of a Diffusing Target in Two and Three Dimensions}

\vspace{3mm}

Ewan McCulloch\textsuperscript{1} and Adam Nahum\textsuperscript{1}

\vspace{2mm}

\textsuperscript{1}\textit{\small Laboratoire de Physique de l'École Normale Supérieure, CNRS,\\
ENS \& Université PSL; 24 rue Lhomond, 75005 Paris, France}

\makeatother

\end{center}

\vspace{0mm}

\begin{center}
\textbf{CONTENTS}
\end{center}

\begingroup
\makeatletter

\def\l@f@section{%
  \addpenalty{\@secpenalty}%
  \addvspace{0.7em}%
}

\@starttoc{apc}

\makeatother
\endgroup

\vspace{0mm}

%\tableofcontents

\section{Continuum descriptions}
\label{sec:continuum_time_descriptions}

In this appendix we derive the continuum KPZ and DPRE descriptions directly from the discrete filtering equation. 
Previously Ref.~\cite{Kim2025}
obtained these continuum formulations by considering a model of  continuous-time Gaussian monitoring: here we retain the discrete
Bernoulli observation model and assume only that the measurement contrast
\(\Delta'\) used by the inferrer is small; the true contrast \(\Delta\)
need not be small.

Ref.~\cite{jin2022kardar} obtained a   modified stochastic heat equation description of monitoring
that is closely related to the KPZ equation in Ref.~\cite{Kim2025} and below but which (interestingly) is different. The KPZ formulation of Ref.~\cite{Kim2025}  
involves a defect at the teacher location $X_t$, and describes the stochastic evolution of the posterior after conditioning on the teacher trajectory. 
By contrast the modified stochastic heat equation in Ref.~\cite{jin2022kardar} makes no reference to the teacher: it describes
the stochastic evolution of the posterior without conditioning on a specific teacher trajectory. The latter formulation is specific to the Bayes-optimal case.\footnote{If we can  use the inferrer's posterior ${P_t = P_t(x)}$ at time $t$ to determine the probabilities of measurement outcomes at time $t$ \cite{jin2022kardar}, then the statistics of ${(P_{t+1}- P_t)}$ is fully determined by $P_t$ itself. 
This is possible in the Bayes-optimal case because in addition to expressing the inferrer's belief, $P_t$ is also the true distribution over teacher locations at time $t$ (conditioned only on earlier measurement outcomes), and therefore determines the probability of measurement outcomes.
(More precisely, we must assume that the distribution from which the initial teacher location $X_0$ is sampled matches the inferrer's initial condition $P_0$.)}

Below we give two independent derivations of continuum descriptions: one in the ``spacetime'' picture, using replicas,  and by taking a controlled continuum limit for the posterior-updating equation.  At the end these are confirmed to be consistent. The explicit values of the couplings $\mathsf g_{ss}$, $\mathsf g_{st}$ may be found in Eqs.~\ref{eq:gst_quadratic},~\ref{eq:gst_kpz_quadratic} below.

\subsection{Directed polymer partition sum}

We begin with the directed-polymer representation of the filtering problem at fixed teacher trajectory \(X_\tau\) (Eq.~\ref{eq:worldinedisorder}, generalized to an arbitrary initial condition):
\begin{equation}
    Z_t(x)
    =
    \sum_{\substack{x_0,\ldots,x_t\\x_t=x}}
    P_0(x_0)
    \prod_{\tau=1}^{t}
    K(x_\tau|x_{\tau-1})
    \exp\!\left[
        \Delta'\eta_\tau(x_\tau)
    \right].
    \label{eq:bernoulli_polymer_partition_sum}
\end{equation}
The candidate trajectory \(\{x_\tau\}\) is therefore a directed polymer in the local spacetime potential \(\Delta'\eta_\tau(x)\).

Introduce \(m\) replicas \(x^a_{\tau}\), \(a=1,\ldots,m\), of the candidate trajectory. These replica candidate trajectories are known as \emph{students}. Ultimately \(m\to0\). Including the target, or \emph{teacher}, trajectory this corresponds to \(n=m+1\) polymers, with the replica limit \(n\to1\). At each spacetime vertex $(x,\tau)$, define
\begin{equation}
    N_\tau(x)
    =
    \sum_{a=1}^{m}\delta_{x_{\tau}^a,x},
    \qquad
    T_\tau(x)
    =
    \delta_{x,X_\tau}.
    \label{eq:student_teacher_vertex_occupations}
\end{equation}
Here \(N_\tau(x)\) counts the number of student polymers at the vertex, while \(T_\tau(x)\in\{0,1\}\) records the presence of the teacher polymer (the target).

Conditioned on the teacher occupation $T=T_{\tau}(x)$, the observation $\eta=\eta_\tau(x)$ at a given space-time vertex $(x,{\tau})$ is Bernoulli distributed,
\begin{equation}
    \Pr(\eta=1\mid T)=p_T,
    \qquad
    \Pr(\eta=0\mid T)=1-p_T,
    \qquad
    p_T=p_0+(p_1-p_0)T.
\end{equation}
Averaging the replicated Boltzmann weight $\exp(\Delta' \eta_\tau(x) N_\tau(x))$ over the observation noise $\eta$ (for a given $T\in \{0,1\}$) therefore gives the exact replica statistical mechanics model Boltzmann factor
\begin{equation}
    W_{\rm B}(N,T)=\mathbb{E}_{\eta|T} \left[e^{\Delta'\eta N}\right]=1-p_T+p_Te^{\Delta'N},
    \label{eq:exact_bernoulli_vertex_rule}
\end{equation}
where the subscript ``$\mathrm{B}$'' 
denotes the Bernoulli measurement model.
Since observations at distinct spacetime points are independent, the averaged replicated partition sum is
\begin{equation}
\begin{split}
    \mathbb{E}_{\eta|X}
        \left[Z^m_t\right]
    ={}&
    \sum_{\{x_{\tau}^a\}}
    \prod_{a=1}^{m}
    \left[
        P_0(x_{0}^a)
        \prod_{\tau=1}^{t}
        K(x_{\tau}^a|x_{\tau-1}^a)
    \right]\times
    \prod_{\tau=1}^{t}\prod_x
    W_{\rm B}\!\left(
        N_\tau(x),T_\tau(x)
    \right),
    \label{eq:replicated_bernoulli_partition_sum}
\end{split}
\end{equation}
where the average is over the noise $\eta$ at fixed target trajectory $X$. We have left the spatial argument(s) of $Z_t$ and the final endpoint conditions on the sums implicit.

A single student in the bulk, i.e., not coincident with the teacher, carries the local weight
\begin{equation}
    W_1
    \equiv
    W_{\rm B}(1,0)
    =
    1-p_0+p_0e^{\Delta'}.
\end{equation}
At every time slice, \(\sum_xN_\tau(x)=m\), so the product of \(W_1^{N_\tau(x)}\) over all vertices is the path-independent factor \(W_1^{mt}\). This factor may therefore be absorbed into the normalization. We define the normalized statistical weight
\begin{equation}
    \widehat W_{\rm B}(N,T)
    =
    \frac{W_{\rm B}(N,T)}{W_1^N}, \qquad \widehat W_{\rm B}(1,0)=1.
    \label{eq:normalized_bernoulli_vertex}
\end{equation}

Writing
\begin{equation}
    \delta p=p_1-p_0,
    \qquad
    v_0=p_0(1-p_0),
    \qquad
    v_1=p_1(1-p_1),
    \qquad
    \delta v=v_1-v_0,
\label{eq:definevdeltap}
\end{equation}
we expand the logarithm of \(\widehat W_{\rm B}(N,T)\) at small \(\Delta'\). Since \(p_T=p_0+\delta p\,T\) and \(p_T(1-p_T)=v_0+\delta v\,T\), we obtain
\begin{equation}
\begin{split}
    \log \widehat W_{\rm B}(N,T)
    ={}&
    \delta p\Delta'\,TN
    +\frac{v_0(\Delta')^2}{2}
    \left(N^2-N\right)
    +\frac{\delta v (\Delta')^2}{2}
   TN^2 +O\!\left((\Delta')^3\right).
\label{eq:bernoulli_vertex_expansion}
\end{split}
\end{equation}
To identify the corresponding replica contacts (which replica polymers are present at the vertex), write
\begin{equation}
    I_a=\delta_{x_{\tau,a},x},
    \qquad
    N=\sum_a I_a.
\end{equation}
Using \(I_a^2=I_a\), we have $N^2-N
    =
    2\sum_{a<b}I_aI_b$ and $TN^2
    =
    T\sum_a I_a
    +
    2T\sum_{a<b}I_aI_b$.
Equation~\eqref{eq:bernoulli_vertex_expansion} can therefore be written as
\begin{equation}
\begin{split}
    \log \widehat W_{\rm B}
    =
    \mathsf g_{ss}
    \sum_{a<b}I_aI_b
    +
    \mathsf g_{st}
    T\sum_a I_a +
   \mathsf g_{sst}
\,    T\sum_{a<b}I_aI_b
    +O\!\left((\Delta')^3\right),
    \label{eq:bernoulli_replica_contacts}
\end{split}
\end{equation}
where
\begin{align}
    \mathsf g_{ss}
    \equiv
    v_0(\Delta')^2,\qquad
    \mathsf g_{st}
    \equiv
    \delta p\,\Delta'
    +\frac{\delta v}{2}(\Delta')^2,
    \qquad 
    \mathsf g_{sst} =
     \delta v\,(\Delta')^2.
    \label{eq:gst_quadratic}
\end{align}
The first term in Eq.~\eqref{eq:bernoulli_replica_contacts} is a contact between two students, while the second is a contact between one student and the teacher. The third term requires two students and the teacher to occupy the same vertex, and is therefore a genuine three-replica contact. 

We now neglect the three-replica contact and the \(O((\Delta')^3)\) terms, retaining only the pairwise vertex weight,
\begin{equation}
    \widehat W_{\rm pair}
    =
    \exp\left[
        \mathsf g_{ss}\sum_{a<b}I_aI_b
        +
        \mathsf g_{st}\,T\sum_a I_a
    \right].
    \label{eq:pairwise_vertex_rule}
\end{equation}
Dropping the $O((\Delta')^3)$ terms is justified straightforwardly by our basic assumption $\Delta'\ll 1$. 
The justification for dropping the three-replica contact uses the RG at intermediate scales and 
is explained  later in this appendix.

We note in passing that the reduced vertex weight (\ref{eq:pairwise_vertex_rule}) is exactly that obtained from a polymer in a spatially homogeneous Gaussian random potential. Let \(V_\tau(x)\) be independent Gaussian variables with
\begin{equation}
    \mathbb{E}_V \left[{V_\tau(x)}\right]=0,
    \qquad
    \mathbb{E}_V \left[V_\tau(x)V_{\tau'}(x')\right] 
    =
    \mathsf g_{ss}\,
    \delta_{\tau,\tau'}\delta_{x,x'},
    \label{eq:gaussian_lattice_potential}
\end{equation}
and supplement the random potential by a deterministic attraction \(\mathsf g_{st}\) to the teacher. The local Boltzmann factor for one student is then $\exp\!\left[V_\tau(x)+\mathsf g_{st}T_\tau(x)\right]$. For \(N\) students at the vertex, averaging over \(V_\tau(x)\) gives
\begin{equation}
    \mathbb{E}_V\left[{
        e^{NV_\tau(x)+N\mathsf g_{st}T}
    }\right]
    =
    \exp\left[
        \frac{\mathsf g_{ss}}{2}N^2
        +\mathsf g_{st}NT
    \right].
\end{equation}
The term \(\mathsf g_{ss}N/2\) is a path-independent one-student
normalization. Removing it gives
\begin{equation}
\begin{split}
    \widehat W_{\rm G}(N,T)
    =
    \exp\left[
        \frac{\mathsf g_{ss}}{2}(N^2-N)
        +\mathsf g_{st}NT
    \right]=
    \exp\left[
        \mathsf g_{ss}\sum_{a<b}I_aI_b
        +
        \mathsf g_{st}T\sum_a I_a
    \right],
    \label{eq:gaussian_replica_vertex}
\end{split}
\end{equation}
which is precisely the vertex weight in
Eq.~\eqref{eq:pairwise_vertex_rule}.

Having obtained (\ref{eq:pairwise_vertex_rule}), the replica polymer partition sum, now also averaged over the target trajectory, now denoted by $x^0_{\tau}\equiv X_\tau$, under the same dynamical prior as the candidate trajectories, is given by
\begin{align}
    \overline{Z_t^{n-1}} 
        = \sum_{x^0_{\tau},\cdots, x^{n-1}_{\tau}} \exp\Bigg\{\sum_{\tau} \Bigg[ \sum_{a=0}^{n-1}\log\left[K(x_{\tau}^a|x_{\tau-1}^a)\right] +\frac{\mathsf g_{ss}}{2}\sum_{\substack{a,b=1 \\ a\neq b}}^{n-1} \delta_{x_{\tau}^a,x_{\tau}^b}
        +
        \mathsf g_{st}\sum_{a=1}^{n-1} \delta_{x_{\tau}^a,x_{\tau}^0}\Bigg]
        \Bigg\},
\end{align}
where boundary conditions are again left implicit. (As in the main text, $\overline{(\bullet)}$ denotes an average over target trajectories and measurement noise $\eta$.)

Small $\Delta'$ implies small contact interaction strengths $\mathsf g_{ss}$ and $\mathsf g_{st}$ between polymers. In this case, even if the polymers bind, they are bound only on a large binding length, and contacts are rare. 
On scales sufficiently smaller than the binding length,
the polymers behave as free random walks. This large lengthscale means that the kernel $K$ can be coarse-grained in a controlled way (i.e. only the long-time, long distance form of the free propagator is important) giving the  following continuum form,
\begin{align}
    \overline{Z_t^{n-1}} 
        \simeq \int \mathcal D x^0(\tau) \cdots \mathcal D x^{n-1}(\tau) \exp\Bigg\{\int d{\tau} \Bigg[ -\sum_{a=0}^{n-1} \frac{(\dot{x}^a)^2}{4D} + \frac{\mathsf g_{ss}}{2}\sum_{\substack{a,b=1 \\ a\neq b}}^{n-1} \delta(x^a(\tau)-x^b(\tau))
        +
        \mathsf g_{st}\sum_{a=1}^{n-1} \delta(x^a(\tau)-x^0(\tau))\Bigg]
        \Bigg\}.
        \label{eq:replica-polymer-path-sum}
\end{align}

This replica action has the symmetry group $S_{n-1}$ associated with the $n-1$ identical student replicas. At Bayes optimality, the assumed and true observation models coincide, $p'_0=p_0$ and $p'_1=p_1$, so that $\Delta'=\Delta$ (Eq.~\eqref{eq:contrasts}). Since the continuum expansion assumes $\Delta'\ll 1$, we may also expand in the small difference $\delta p \equiv p_1-p_0$. Taylor expanding the definition of the true contrast about $p_1=p_0$ gives
\begin{align}
    \Delta
    =
    \log\!\left[
        \frac{p_1/(1-p_1)}{p_0/(1-p_0)}
    \right]=
    \frac{\delta p}{v_0}
    +
    \frac{2p_0-1}{2v_0^2}\,\delta p^2
    +
    O(\delta p^3).
\end{align}
Inverting this expansion gives
\begin{equation}
    \delta p
    =
    v_0\Delta
    +
    \frac{v_0(1-2p_0)}{2}\Delta^2
    +
    O(\Delta^3).
\end{equation}
Moreover, since $\delta v={\mathcal O}(\Delta)$, substituting $\Delta'=\Delta$ into Eq.~\eqref{eq:gst_quadratic} gives
\begin{align}
    \mathsf g_{st}=\delta p\,\Delta+\frac{\delta v}{2}\Delta^2=\mathsf g_{ss}+O(\Delta^3).
\end{align}
Thus, to the quadratic order, $\mathsf g_{st}=\mathsf g_{ss}$. Since the teacher and students also have the same dynamical prior, the $S_{n-1}$ symmetry among the students is enhanced to full $S_n$ permutation symmetry.

\subsection{KPZ equation}

Next we consider the KPZ formulation (which does not use replicas).
Again we may  neglect the third and higher cumulants of $\eta$ (see below), which is equivalent to replacing $\eta$ with a Gaussian variable. This corresponds to the lattice transfer equation
\begin{equation}
    Z_{\tau+1}(x)
    =
    \exp\!\left[
        V_{\tau+1}(x)
        +\mathsf g_{st}\delta_{x,X_{\tau+1}}
    \right]
    \sum_y K(x|y)Z_\tau(y),
    \label{eq:gaussian_lattice_transfer}
\end{equation}
where the random potential $V$ is Gaussian noise satisfying Eq.~\eqref{eq:gaussian_lattice_potential}. Equation~\eqref{eq:gaussian_lattice_transfer} is equivalently the discrete-time stochastic heat equation associated with the Gaussian directed polymer, with a pin along $X_\tau$. Writing ${h_\tau(x)=\log Z_\tau(x)}$ gives the exact discrete height-field equation
\begin{equation}
\begin{split}
    h_{\tau+1}(x)-h_\tau(x)
    =
    V_{\tau+1}(x)
    +\mathsf g_{st}\delta_{x,X_{\tau+1}}+
    \log\!\left[
        \frac{
            \sum_y K(x|y)e^{h_\tau(y)}
        }{
            e^{h_\tau(x)}
        }
    \right].
    \label{eq:gaussian_discrete_height}
\end{split}
\end{equation}
Thus the random potential $V$ enters additively in the height-field equation.

For small \(\Delta'\), the random potential and the pinning interaction are weak: $\mathsf g_{ss}=O\!\left((\Delta')^2\right)$ and $\mathsf g_{st}=O(\Delta')$. There is consequently a large separation between the lattice scale and the scales on which the coarse-grained disorder is (potentially) $\mathcal O(1)$. On shorter scales, the dynamics is diffusive, and so the height field is expected to be smooth and slowly varying on the microscopic scale. We therefore expanded the final term in Eq.~\eqref{eq:gaussian_discrete_height} in spatial gradients, and the initial term in time derivatives. Note that it is not necessary to rescale the lattice spacing, which is fixed to~1.

For an isotropic transition kernel, this expansion gives
\begin{equation}
    h_{\tau+1}(x)-h_\tau(x) = \partial_\tau h + \cdots,  \qquad
    \log\!\left[
        \frac{
            \sum_y K(x|y)e^{h_\tau(y)}
        }{
            e^{h_\tau(x)}
        }
    \right]
    =
    D\nabla^2h
    +D(\nabla h)^2+
    \cdots,
    \label{eq:gaussian_height_gradient_expansion}
\end{equation}
where \(D\) is the diffusion constant of the underlying random walk, and where $\cdots$ denotes higher derivative/gradient terms. Under coarse graining, the independent lattice Gaussian variables \(V_\tau(x)\) flow to the same Gaussian white-noise field as a continuum spacetime-white potential with the corresponding variance:
\begin{equation}
    \overline{\xi(x,t)\xi(x',t')}
    =
    \mathsf g_{ss}\,
    \delta^{(d)}(x-x')\delta(t-t').
    \label{eq:continuum_noise_covariance}
\end{equation}

The resulting continuum equation is
\begin{equation}
    \partial_t h
    =
    D\nabla^2h
    +D(\nabla h)^2
    +\xi(x,t)
    +\mathsf g_{st}\delta^{(d)}(x-X_t).
\label{eq:kpz_from_gaussian_polymer}
\end{equation}
This is the KPZ equation with a diffusing attractive defect, with couplings given to quadratic order in $\Delta'$ by
\begin{align}
    \mathsf g_{ss}
    &=
    p_0(1-p_0)(\Delta')^2
    +O\!\left((\Delta')^3\right),
    \label{eq:gss_kpz_quadratic}
\\
    \mathsf g_{st}
    &=
    (p_1-p_0)\Delta'
    +
    \frac{1}{2}
    \left[
        p_1(1-p_1)-p_0(1-p_0)
    \right](\Delta')^2
    +O\!\left((\Delta')^3\right).
    \label{eq:gst_kpz_quadratic}
\end{align}
Equivalently, Eq.~\eqref{eq:kpz_from_gaussian_polymer} describes the free
energy of the continuum directed polymer
\begin{equation}
\begin{split}
    Z(x,t)
    =
    \int_{x(t)=x}\mathcal D x(\tau)\,
    \exp\Bigg[
        -\int_0^t d\tau\,
        \frac{\dot{x}(\tau)^2}{4D}
        +\int_0^t d\tau\,\xi(x(\tau),\tau)+\mathsf g_{st}
        \int_0^t d\tau\,
        \delta^{(d)}\!\bigl(x(\tau)-X(\tau)\bigr)
    \Bigg].
    \label{eq:continuum_gaussian_polymer}
\end{split}
\end{equation}
Replicating this partition sum, averaging over the noise $\xi$ and the target trajectory $X$, yields the replica polymer path sum in Eq.~\eqref{eq:replica-polymer-path-sum}.

\subsection{Higher cumulants}

The Bernoulli and Gaussian 
measurement protocols agree in their first and second moments, or equivalently, agree
for every local
configuration involving at most two replica polymers in the replica description. 
They therefore give the same pairwise student--student and student--teacher couplings $\mathsf g_{ss}$ and $\mathsf g_{st}$. Their first difference occurs for configurations involving three or more polymers. For example, the third term in Eq.~\eqref{eq:bernoulli_replica_contacts} is a genuine three-replica interaction involving two students and the teacher. 
Further higher-replica interactions are contained in the higher orders of the small-$\Delta'$ expansion.
The three-replica interaction is initially at the same scale (order $\Delta'^2$) as the two-replica interactions: here we explain why it can nevertheless be dropped
in the regime where the derivation is quantitatively controlled, i.e. at small $\Delta'$. 

Note that this implies only that the continuum theory with 2-replica interactions is a quantitatively accurate \textit{effective field theory}, i.e. with a UV cutoff lengthscale (which can be taken to be of the order of the  lattice spacing).
This effective field theory is quantitatively useful for various calculations that can be performed by weak-coupling RG. 
Our statement does not imply that 
a description with only two-replica interactions always remains valid  under RG in which the cutoff is increased to an arbitrarily large scale. Whether this is the case or not (for small bare $\Delta'$) depends on the dimensionality.

For small $\Delta'$ we assume that perturbative RG is valid at least in the early stages of RG when all couplings are small (in $d>2$ this remains true on all scales).
Higher-replica interactions
may be generated from pairwise couplings:
several pairwise collisions occurring within the same coarse-grained spacetime block produce an effective local interaction among three or more replicas. 
The RG flow is, however, triangular in replica number. Pairwise couplings renormalize autonomously, whereas a $k$-replica coupling can depend on couplings involving at most $k$ replicas. 
Consequently, the different microscopic higher-replica vertices in the Bernoulli and Gaussian models do not affect the RG flows of $\mathsf g_{ss}$ and $\mathsf g_{st}$. 

We can also show that, in the weak-measurement regime,
the differences between the protocols through the values of the
higher-replica  couplings
is rapidly suppressed, so that the protocols become essentially equivalent after coarse graining to some scale $\ell$, where $\ell\gg 1$.

A local contact involving $k$ polymers has the form
\begin{equation}
    \mathcal O_k
    =
    \int dt\,
    \prod_{a=2}^{k}
    \delta^{(d)}
    \bigl(\boldsymbol x^a(t)-\boldsymbol x^1(t)\bigr).
    \label{eq:k_replica_contact}
\end{equation}
At the EW fixed point, under $\boldsymbol x\rightarrow b\boldsymbol x$ and $t\rightarrow b^2t$, this operator has RG eigenvalue $y_k=2-(k-1)d$. Thus higher-replica contacts are increasingly irrelevant with increasing $k$. In particular, $y_2=2-d$ and $y_3=2-2d$.

To make the suppression explicit, let $u_2$ denote a pairwise coupling and $u_3$ a three-replica coupling. In $d=3$, their weak-coupling flows have the schematic form
\begin{equation}
    \frac{du_2}{dl}
    =
    -u_2+\cdots,
    \qquad
    \frac{du_3}{dl}
    =
    -4u_3+B u_2^2+\cdots.
    \label{eq:three_replica_flow_d3}
\end{equation}
The term proportional to $u_2^2$ is the three-replica interaction generated by successive pairwise collisions. Because the pairwise couplings are identical in the Bernoulli and Gaussian theories, this source term is also identical. Their difference $\delta u_3\equiv u_3^{\rm B}-u_3^{\rm G}$ therefore decays under perturbative RG, $\delta u_3(l) \simeq e^{-4l}\delta u_3(0)$.

In $d=2$, the pairwise interaction is marginally relevant while the three-replica interaction remains irrelevant:
\begin{equation}
    \frac{du_2}{dl}
    =
    A u_2^2+\cdots,
    \qquad
    \frac{du_3}{dl}
    =
    -2u_3+B u_2^2+\cdots.
    \label{eq:three_replica_flow_d2}
\end{equation}
Consider a small bare pairwise coupling. The theory remains well-described by the perturbative RG until the RG time $l_*\sim({A u_2(0)})^{-1}$, corresponding to the exponentially large crossover length $L_*\sim e^{l_*}$. Using again the fact that the two-replica couplings are identical for the Gaussian and Bernoulli problems, the difference $\delta u_3\equiv u_3^{\rm B}-u_3^{\rm G}$ in the three-replica couplings obeys
\begin{equation}
    \frac{d\,\delta u_3}{dl}
    =
    \bigl[-2+O(u_2)\bigr]\delta u_3.
\end{equation}
Under perturbative RG up to the crossover scale, it thus decays to $\delta u_3(l_*)\sim\delta u_3(0)/L_*$.

Contacts involving more replicas are still more
irrelevant. It is consequently consistent, at weak measurement
strength, to omit the bare higher-replica interactions.

\section{Replica polymers and one-loop flow}
\label{app:replica-boson-RG}

In this appendix we derive the weak-coupling flow quoted in the main text. We start from the continuum replica directed-polymer partition function in Eq.~\eqref{eq:replica-polymer-path-sum}. This partition sum can equivalently be written in terms of a replica polymer action $S_{n}$ for $n-1$ student replicas, $a=1,\cdots,n-1$, and one teacher replica, $a=0$,
\begin{equation}
S_n[\{x^a(\tau)\}]=\int d\tau \Bigg[\sum_{a=0}^{n-1} \frac{(\dot{x}^a)^2}{4D} - \frac{\mathsf g_{ss}}{2}\sum_{\substack{a,b=1 \\ a\neq b}}^{n-1} \delta(x^a(\tau)-x^b(\tau))
- \mathsf g_{st}\sum_{a=1}^{n-1} \delta(x^a(\tau)-x^0(\tau))\Bigg].
\label{eq:polymer-path-integral}
\end{equation}

\subsection{Replica boson action}

Define the coupling matrix
\begin{equation}
\mathsf g_{ab}
=
\begin{cases}
\mathsf g_{ss},
& a,b\geq 1,\quad a\neq b,\\
\mathsf g_{st},
& a=0,\ b\geq1\ \text{or}\ b=0,\ a\geq1,\\
0,
& a=b.
\end{cases}
\label{eq:replica-coupling-matrix}
\end{equation}
The path integral in Eq.~\eqref{eq:replica-polymer-path-sum} can be equivalently written in terms of the first-quantized Hamiltonian
\begin{equation}
H_n
=
-D\sum_{a=0}^{n-1}\nabla_a^2
-\frac12
\sum_{\substack{a,b=0\\a\neq b}}^{n-1}
\mathsf g_{ab}\,
\delta^{(d)}
\bigl({x}^a-{x}^b\bigr).
\label{eq:first-quantized-replica-H}
\end{equation}
For fixed initial and final polymer positions,
\begin{equation}
Z_n
\bigl(\{{x}_{f}^{a}\},t
\mid
\{{x}_{i}^{a}\},0\bigr)
=
\left\langle
\{{x}_{f}^{a}\}
\right|
e^{-t H_n}
\left|
\{{x}_{i}^{a}\}
\right\rangle .
\label{eq:replica-propagator}
\end{equation}
This transfer-matrix problem has previously been mapped, by second quantization, to a theory of complex bosons~\cite{KARDAR1987582,1992PhRvA..45.8727K,
PhysRevB.39.9153,PhysRevB.42.10113}. In this formulation the
corresponding grand-canonical partition function is written as the coherent-state functional integral
\begin{equation}
    \mathcal Z_n^{\mathrm{gc}}
    =
    \int
    \prod_{a=0}^{n-1}
    \mathcal D\psi_a\,\mathcal D\bar\psi_a\,
    \exp\!\left(-S_n[\psi,\bar\psi]\right),
    \label{eq:grand-canonical-boson-functional}
\end{equation}
with
\begin{equation}
S_n[\psi,\bar\psi]
=
\int dt\,d^d\boldsymbol{x}\,
\Bigg[
\sum_{a=0}^{n-1}
\bar\psi_a
(\partial_t-D\nabla^2)\psi_a
-\frac12
\sum_{\substack{a,b=0\\a\neq b}}^{n-1}
\mathsf g_{ab}\,
\bar\psi_a\psi_a
\bar\psi_b\psi_b
\Bigg].
\label{eq:boson_action_general}
\end{equation}
Here $\psi_a$ and $\bar\psi_a$ are complex
bosonic fields.

This is not identical to the original replica-polymer partition function: the latter belongs to the sector containing exactly one particle of each replica species, whereas $\mathcal Z_n^{\mathrm{gc}}$ sums over boson-number sectors. However, this distinction is not important for the bulk renormalization of the two-body contact couplings $\mathsf g_{ab}$, whose RG flows may be computed using the boson action $S_n[\psi,\bar\psi]$.

\subsection{One-loop beta functions}

The continuum 
contact interaction in (\ref{eq:boson_action_general}) must be defined with a short-distance regulator, since loops involving coincident contacts are logarithmically UV divergent at \(d=2\) and power-law divergent for \(d>2\). We use a momentum cutoff \(\Lambda\), representing the inverse microscopic length scale; for an underlying lattice spacing \(a\), \(\Lambda=c_\Lambda/a\) with nonuniversal \(c_\Lambda=O(1)\).

We define the beta function ${\dd g_{ab}/\dd l}$ by integrating out modes in a thin momentum shell \(\Lambda e^{-\dd l}<|q|<\Lambda\), followed by the diffusive rescaling \(x\to e^{\dd l}x\), \(t\to e^{2\dd l}t\) that restores the cutoff.
Power counting gives the tree-level contribution \((2-d)g_{ab}\) to the beta function, since a contact interaction has engineering dimension \(2-d\). The one-loop correction comes from two successive \(ab\) contacts.  The free boson propagator is
\[
G_0(q,\omega)\equiv \langle \psi_a(q,\omega)\bar\psi_b(-q,-\omega)\rangle_0
=
\delta_{ab}\big[-i\omega+Dq^2\big]^{-1},
\]
with the pole prescription understood so that propagation is retarded. Thus the repeated \(ab\) bubble gives
\[
\delta \mathsf g_{ab}
=
\mathsf g_{ab}^2
\int_{\rm shell}\frac{\dd^dq}{(2\pi)^d}
\int\frac{\dd\omega}{2\pi}\,
G_0(q,\omega)G_0(-q,-\omega).
\]
The frequency integral gives
\[
\int\frac{\dd\omega}{2\pi}
\frac{1}{\omega^2+D^2q^4}
=
\frac{1}{2Dq^2}.
\]
Therefore
\[
\delta \mathsf g_{ab}
=
\mathsf g_{ab}^2
\frac{K_d}{2D}\Lambda^{d-2}d l,
\quad
K_d=\frac{1}{2^{d-1}\pi^{d/2}\Gamma(d/2)}.
\]
Combining this with the tree-level scaling gives the dimensionful flow
\begin{equation}
\frac{d \mathsf g_{ab}}{d l}
=
(2-d)\mathsf g_{ab}
+
\frac{K_d}{2D}\Lambda^{d-2}\mathsf g_{ab}^2
+
O(\mathsf g_{ab}^3).
\end{equation}
Equivalently, for the dimensionless coupling $g_{ab}=\frac{K_d}{2D}\Lambda^{d-2}\mathsf g_{ab}$, we obtain
\begin{equation}
\label{eq:apprescaledrgeqn}
\frac{d g_{ab}}{d l}
=
(2-d) g_{ab}
+ g_{ab}^{\,2}.
\end{equation}
The calculation above establishes the flow to quadratic order. In fact, because the propagators are retarded, the only nonvanishing vertex corrections are repeated $(ab)$ bubbles. These generate no independent higher-order terms in the beta function, which is therefore one-loop exact~\cite{Wiese1998}.

The weak-coupling growth of the localization length in \(d=2\) is fixed by the marginal flow.  At Bayes optimality, \(\mathsf g_{ss}=\mathsf g_{st}\equiv \mathsf g\), and the dimensionless coupling is $g={\mathsf g/(4\pi D)}$, where we have used \(\Lambda^{d-2}=1\) for \(d=2\) and \(K_2=1/(2\pi)\). Integrating \(dg'/d l=g'^2\) up to any fixed \(O(1)\) reference value \(g_*\) gives
\[
l_*=\int_{g}^{g_*} dg'\, g'^{-2}
=
g^{-1}-g_*^{-1}
=
4\pi D\,\mathsf g^{-1}+O(1).
\]
Here \(l_*\) is the RG time at which the contact interaction reaches strong coupling.  Since, in the momentum-shell RG defined above, \(e^{l_*}\) is the corresponding spatial rescaling factor, the physical RMS localization length is $\ell_{\rm rms}\sim e^{l_*}$ in microscopic length units.  Equivalently,
\begin{equation}
\log \ell_{\rm rms}
=
4\pi D\,\mathsf g^{-1}+O(1) \sim
\frac{16\pi D}{\Delta^2},
\label{eq:rmsapp}
\end{equation}
where we have used the weak-measurement matching \(\mathsf g=\Delta^2/4\), valid for $\mathsf g \ll 1$, in the second equality.

\subsection{Misspecification of the diffusion constant}

The above generalizes immediately to the case where the teacher and students have different diffusion constants $D_t$ and $D_s$ (denoted $D$ and $D'$ in the main text), or more generally where each replica $a$ has its own diffusion constant $D_a$. The factor of $1/(2D)$ in the loop integral for $\delta \mathsf g_{ab}$ becomes $1/(D_a + D_b)$.\footnote{Physically the replacement of $2D$ with  $D_{a}+D_b$ is just because $D_a+D_b$ is the diffusion constant of the \textit{relative} coordinate in the two-particle problem.}
After redefining 
$g_{ab}=\frac{K_d}{{D_a+ D_b}}\Lambda^{d-2}\mathsf g_{ab}$,
the RG equations again take the form in Eq.~\ref{eq:apprescaledrgeqn}.
Therefore, in 2+1D and in the limit of weak coupling, the line
\begin{equation}
\mathsf g_{st}  =
\frac{D_s + D_t}{2 D_s} 
\mathsf g_{ss}
\label{app:optimalproxy}
\end{equation}
in the plane of bare coupling constants ${(\mathsf g_{ss}, \mathsf g_{st})}$ 
plays a similar role to the Bayes-optimal line of the $D_s=D_t$ case, in that,
in 2+1D, the boundary between the two phases  should be asymptotically tangent to the line (\ref{app:optimalproxy}) at small $(\mathsf g_{ss}, \mathsf g_{st})$.

It is also interesting to consider the tracking phase diagram as a function of $D_s$. 
For concreteness, consider the 2+1D case, and  set  ${\mathsf g_{st} = \mathsf g_{ss}=\mathsf g}$
so that the only source of non-optimality is  mischaracterization of the diffusion constant.
A simple extension of the analysis of the phase boundary in the main text shows that 
\begin{equation}
    D_s^c = D_t  \left( 1 - \frac{\mathsf g \lambda_c}{2\pi} + \ldots \right)
\end{equation}
at small $\mathsf g$, where $\lambda_c$ is a constant,
with the successful (pinned) phase at $D_s>D_s^c$.
The critical point on the $D_s$ axis is slightly below the Bayes-optimal point $D_s=D_t$.

Generalizing (\ref{eq:rmsapp}), the flow of $\mathsf g_{st}$ shows that the 
bound state size 
 grows as 
 \begin{equation}
     \ell_{\mathrm rms}\sim\exp\left( \frac{2\pi (D_s + D_t)}{\mathsf g_{st}} \right)
      \end{equation}
when $D_s$ is increased. 
This formula is valid so long as we are not too close to the phase transition (where $\mathsf g_{ss}$ cannot be neglected).
Since $\ell_{\mathrm rms}$ must also diverge when $D_s \to D_s^c$ from above, 
we see that $\ell_{\mathrm rms}$ depends non-monotonically  on $D_s$ within the pinned phase.

{

\subsection{Generalization of the $\epsilon$ expansion for $k$-replica binding}
\label{app:kreplica}

In Sec.~\ref{sec:rgcomments} we claim that the phase transition for $\overline{\mathcal{Z}^k}$ (for the DPRE, with ${k=2,3,4,\ldots}$) is in fact governed by a nontrivial fixed point for the $k$-replica coupling $g_k$, with lower-replica couplings being irrelevant. 
A  $k$-replica interaction is one that is nonzero only when $k$ replicas 
(or $k$ walks, or $k$ quantum particles, depending on the language we use) approach each other on the scale of the cutoff.

Many points of the argument below will be heuristic.

First let us distinguish between the \textit{bare} coupling constant $g$, which describes bare two-replica  interactions, and \textit{renormalized} coupling constants $g_k(\ell)$ at RG time $\ell$.
In the bare theory (resembling Eq.~\ref{eq:replica_polymer_action} but without the ``teacher''), we have  ${g_2(0) = g}$, and ${g_k(0)=0}$ for ${k>2}$.
After renormalization, however, we may develop higher-replica interactions.

To be concrete, we may imagine defining the theory with a UV cutoff. In this case the above couplings are somewhat schematic, because a priori  the $k$-replica interactions may require multiple coupling constants ${\mathbf{g}_k = (g_k^{(1)}, g_k^{(2)}, \ldots)}$ to parameterize them. However, this does not change the key point below.

Recall that the key feature of the RG flows is that the beta function for $g_l(\ell)$ is independent of $g_{l'}(\ell)$ for all $l'>l$. 
This triangular structure can for example be seen diagrammatically
(e.g. there is no way for a 3-point vertex to renormalize a 2-point vertex).
It implies that the RG flows of $g_l(\ell)$ are identical, regardless of whether we study   $\overline{\mathcal{Z}^l}$, or whether we study 
$\overline{\mathcal{Z}^{l'}}$ for some ${l'>l}$.

We assume that the phase transition for $\overline{\mathcal{Z}^k}$  is a binding transition for $k$ walks (equivalently, $k$ quantum particles), with a trivial (free) unbound phase for ${g<g_c^{(k)}}$, a bound phase for ${g>g_c^{(k)}}$, and a scale-invariant critical point at ${g=g_c^{(k)}}$.

Importantly, we expect that the value of $g_c^{(k)}$ is  strictly decreasing in $k$ for $k=2,3,4,\ldots$. This is consistent with the ``Efimov effect'' (three particles can be bound without two particles being bound~\cite{efimov1970,naidon2017efimov}) and is also clearly the case at large $k$, since the binding energy (${\propto k^2}$) scales much faster than entropy ($\propto k$). It is also consistent with rigorous results~\cite{PhysRevLett.73.1464}.
As a result, at the bare coupling value $g=g_c^{(k)}$ at which $\overline{\mathcal{Z}^k}$ is critical, the ``lower'' partition functions  $\overline{\mathcal{Z}^l}$ (${l<k}$) lie inside the trivial unbound phase.

Therefore, at the $k$-replica critical point  $g=g^{(k)}_c$, the $l$-replica couplings flow asymptotically to zero for all ${l<k}$.
If this was not the case, then $\overline{\mathcal{Z}^l}$ would not be in the weak-coupling phase at $g_c^{(k)}$. 

Therefore, in order to study the critical properties of the $k$-walk ($k$-boson) partition function $\overline{\mathcal{Z}}$, it is sufficient to study an effective field theory with only $k$-replica interactions.

Given this fact, it is reasonable to consider an $\epsilon$ expansion for a theory with only $k$-replica interactions, which we could write schematically either using paths $x_a(t)$ with multi-delta-function interactions, or in second quantized form as
\begin{equation}
\mathcal{L}_k= \sum_{a=1}^k
\bar\psi_a
(\partial_t-D\nabla^2)\psi_a
-g_k
\hspace{-2mm}
\sum_{a_1<\cdots<a_k}\hspace{-2mm}
(\bar \psi_{a_1}\psi_{a_1})\cdots (\bar \psi_{a_k}\psi_{a_k}).
\label{eq:krepbosons}
\end{equation}
Power counting gives the critical dimensionality 
\begin{equation}
d_k = \frac{2}{k-1}
\end{equation}
about which the epsilon expansion is performed.
(Note that, although lower-replica interactions are more relevant at tree level than $g_k$, the triangularity mentioned above means they cannot be generated under the RG, regardless of the RG scheme.)

Recall that in the 2-replica problem the beta function truncated at quadratic order, because the allowed diagrams were simply sequences of vertices connected by the square $G(x,t)^2$ of the propagator (if we work in real space).
Here we have the same structure: the allowed diagrams are sequences of the $k$-replica vertex, connected by $G(x,t)^k$.
So we expect that the RG equation again truncates:
\begin{equation}
{d g_k}/{d\ell}
=
y_k g_k+ C_k g_k^{\,2},
\label{eq:supp-generic_beta}
\end{equation}
where $y_k= 2-(k-1)d$ and $C_k$ is a constant.

Therefore, continuing to physical integer dimensionalities $d>2$, we expect a (spatial) correlation length exponent
\begin{equation}\label{eq:nuk}
\nu_k  =  \frac{1}{(k-1)d - 2}.
\end{equation}
for the binding of $k$ replicas. The correlation time $\nu_\parallel = z \nu$ is related to $\nu\equiv \nu_\perp$ by the dynamical exponent $z=2$.
It would be interesting to test (\ref{eq:nuk}) in numerical simulations of $k$ walks/particles.

We might conjecture that $\nu_k$ can be continued to some range of $k\in \mathbb{R}$.  As noted in the text, $\nu_k$ diverges as $k\to 1+2/d$, suggesting (in retrospect) the possibility of the infinite-order transition for the directed polymer problem ($k\to 0$) that has been rigorously demonstrated in the mathematical literature.
}

\section{Numerical simulations}
\label{app:numerics}

We simulate the real-time tracking posterior dynamics defined in Eq.~\eqref{eq:Zupdate} on a periodic $L^d$ lattice, and construct the posterior at an intermediate time using Eq.~\eqref{eq:two_sided_posterior}. For each simulation, a target trajectory and measurement record are generated using the true parameters $(D,p_0,p_1)$, while the observer evolves the posterior using the assumed parameters $(D,p'_0,p'_1)$. 
After discarding an initial transient, we measure the RMS
distance from the true target position $\ell_{\rm rms}$ (Eq.~\ref{eq:rmslength}), and the contact probability $\mathcal P(0)\equiv \mathcal P_\tau({X}_\tau)$, written here in the target-centered frame. Expectation values below include
both a time average in the stationary regime and an average over independent
realizations of the target trajectory and measurement noise.

\subsection{Retrospective inference}
All numerical results, apart from the posterior snapshots shown in Fig.~\ref{fig:2D_main} and Fig.~\ref{fig:3D_main}, use retrospective tracking. The snapshots instead use real-time tracking. Since retrospective tracking is equivalent to performing real-time tracking on the past measurement record and on the time-reversed future measurement record, and then taking a sitewise product before normalizing, the two problems have the same phase diagram and the same bulk universality classes. In particular, correlation-length and dynamical exponents are unchanged. 

However, real-time and retrospective tracking probe different locations along the candidate polymer (candidate worldline): the real-time posterior is an endpoint observable, whereas the retrospective posterior is evaluated at  time in between the initial and final endpoints, i.e. in the temporal ``bulk''. Exponents associated with boundary observables can differ from their bulk counterparts. 
In particular, the contact fraction \(P_t(X_t)\) is the probability of contact at the temporal boundary, while \(\mathcal P_\tau(X_\tau)\), for \(0\ll\tau\ll t\), measures contacts in the (temporal) bulk. 
Since most results in the polymer-pinning literature concern such bulk observables, we primarily use retrospective tracking when extracting critical properties. However,  the posterior snapshots shown in the main text are obtained from real-time tracking.

For retrospective tracking, the target position at a given time is inferred using a long measurement history in both the past and future. The corresponding two-sided posterior is, up to normalization, given by~\cite{Kitagawa1994}
\begin{equation}
    \mathcal P_\tau({x})
    \propto
    P^{\rightarrow}_\tau({x})P^{\leftarrow}_{t-\tau}({x}),
    \label{eq:two-sided-posterior}
\end{equation}
where $P_\tau^{\rightarrow}$ is obtained by forward evolving an initial prior.
For the present system, with time-reversal-symmetric target dynamics and independent measurement noise, $P_{t-\tau}^{\leftarrow}$ is obtained in an equivalent way, as discussed in Sec.~\ref{sec:model}.
Therefore we construct  a single-time sample of the retrospective tracking posterior
$\mathcal P_\tau$  from two independent stationary real-time tracking runs, starting from a flat prior.
One translates the posteriors so that the true target positions coincide at the intermediate time,  multiplies them pointwise, and normalizes. 
Using translational symmetry,
this is equivalent to the ``physical'' retrospective tracking problem in the case where our prior is flat.

This procedure samples equal-time observables of the two-sided posterior in Eq.~\eqref{eq:two-sided-posterior} (although not their correlations at different times). We use it to  sample  the RMS length and the contact fraction. 
For sufficiently long measurement records and large enough $\tau$, $t-\tau$ the initial and final boundary conditions are forgotten and their choice does not affect the stationary expectation values of $\ell_{\rm rms}$ and $\mathcal P(0)$. 
We run  the forward and backward  dynamics for a long  time $\tau = t-\tau = t_{\rm burn}$ (see below) before taking data in order to avoid boundary effects.

\subsection{\texorpdfstring{$3+1$D}{3+1D} depinning transitions}
\label{app:3d-numerics}

We consider the three $d=3$ scans shown in the main text. For the EW--to--pinned and KPZ--to--pinned transitions, we take vertical cuts through the $(g_{\rm ss},g_{\rm st})$ phase diagram by varying the true measurement contrast $\Delta$ at fixed assumed model. The assumed parameters are $(p'_0,p'_1)=(0.10,0.214)$ for the EW scan and $(p'_0,p'_1)=(0.5,0.95)$ for the KPZ scan, with $p_0=p'_0$ in each case. For the Bayes-optimal scan, we set $p_0=p'_0=0.5$ and vary $p_1=p'_1$, so that $\Delta=\Delta'$ throughout. All three scans use $D\simeq 0.0333$. We use approximately $10^6$ stationary samples for each pair $(L,\Delta)$.

We use an initial simulation time $t_{\rm burn}=10L^2$ for all three scans.\footnote{In the $d=2$ simulations, we use $t_{\rm burn} =  10 L^2$. The spectral gap in two dimensions is $\gamma_L^{-1} \simeq L^2/(4\pi^2 D) \simeq 0.51\,L^2$ for $D\simeq 0.05$, as used in the KPZ-pinned scans in Fig.~\ref{fig:2D_main}. The burn-in time is thus about 20 times the relaxation time of the slowest non-uniform diffusive mode, such that short time transients can be ignored.}
To put this timescale in context, the spectral gap of the free random walk on the periodic $L^3$ lattice is $\gamma_L=4D\sin^2(\pi/L)\simeq4\pi^2D/L^2$. For $D\simeq0.0333$, the corresponding diffusive relaxation time is $\gamma_L^{-1}\simeq0.76L^2$, so that $t_{\rm burn}\approx13\gamma_L^{-1}$. The slowest nonuniform mode of a free diffusive process is suppressed by a factor of approximately $e^{-13}\simeq2\times10^{-6}$ during the initial simulation time. 
This strongly suggests that in the cases with $z=2$ dynamics transient effects will be very small.

As a direct convergence check, Fig.~\ref{fig:3d-burnin} compares the EW data with an independently generated dataset using $t_{\rm burn}\approx40\gamma_L^{-1}$. The two agree very well across the depinning transition, with no systematic change in the finite-size curves. 

For the KPZ scan, the delocalized posterior wanders superdiffusively: its finite-size relaxation time scales as $L^{z_{\rm KPZ}}$ with $z_{\rm KPZ}<2$. The choice $t_{\rm burn}=10L^2$ is therefore parametrically longer than this relaxation time at large $L$.

\begin{figure}[t!]
    \centering
    \includegraphics[width=0.4\linewidth]
        {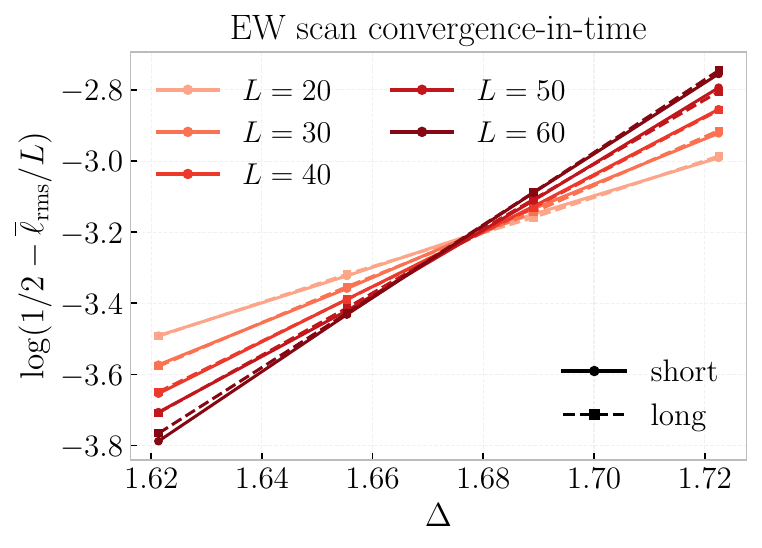}
    \caption{
        Convergence in simulation time for the EW--to--pinned scan.
        Solid lines show the data obtained with (short) initial
        equilibration time $t_{\rm burn}\approx13\gamma_L^{-1}$, while dashed lines show data obtained with the (long) initial equilibration time $t_{\rm burn}\approx40\gamma_L^{-1}$. Here $\gamma_L\simeq4\pi^2D/L^2$ is the spectral gap for a free random walk in three dimensions for system width $L$.
    }
    \label{fig:3d-burnin}
\end{figure}

Now let us turn to the critical properties.
Figure~\ref{fig:3d-raw-width} shows the RMS length $\overline{\ell}_{\rm rms}$. In the unpinned phases, the attraction to the pin flows to zero, and translational invariance gives $\overline{\ell}_{\rm rms}/L\rightarrow 1/2$ (we define the distance from the true target position as the minimum distance on the periodic lattice). Consequently, curves of $\overline{\ell}_{\rm rms}/L$ become strongly compressed near $1/2$. This is shown in Fig.~\ref{fig:3d-raw-width} (upper). We therefore analyze instead
\begin{equation}
    Y_L(\Delta)
    \equiv
    \log\left(
        {1}/{2}
        -
        {\overline{\ell}_{\rm rms}}/{L}
    \right).
    \label{eq:YL-definition}
\end{equation}
This is shown in Fig.~\ref{fig:3d-raw-width} (lower), where the curves appear to clearly cross at the depinning transitions.

\begin{figure}[t!]
    \centering
    \includegraphics[width=0.71\linewidth]
        {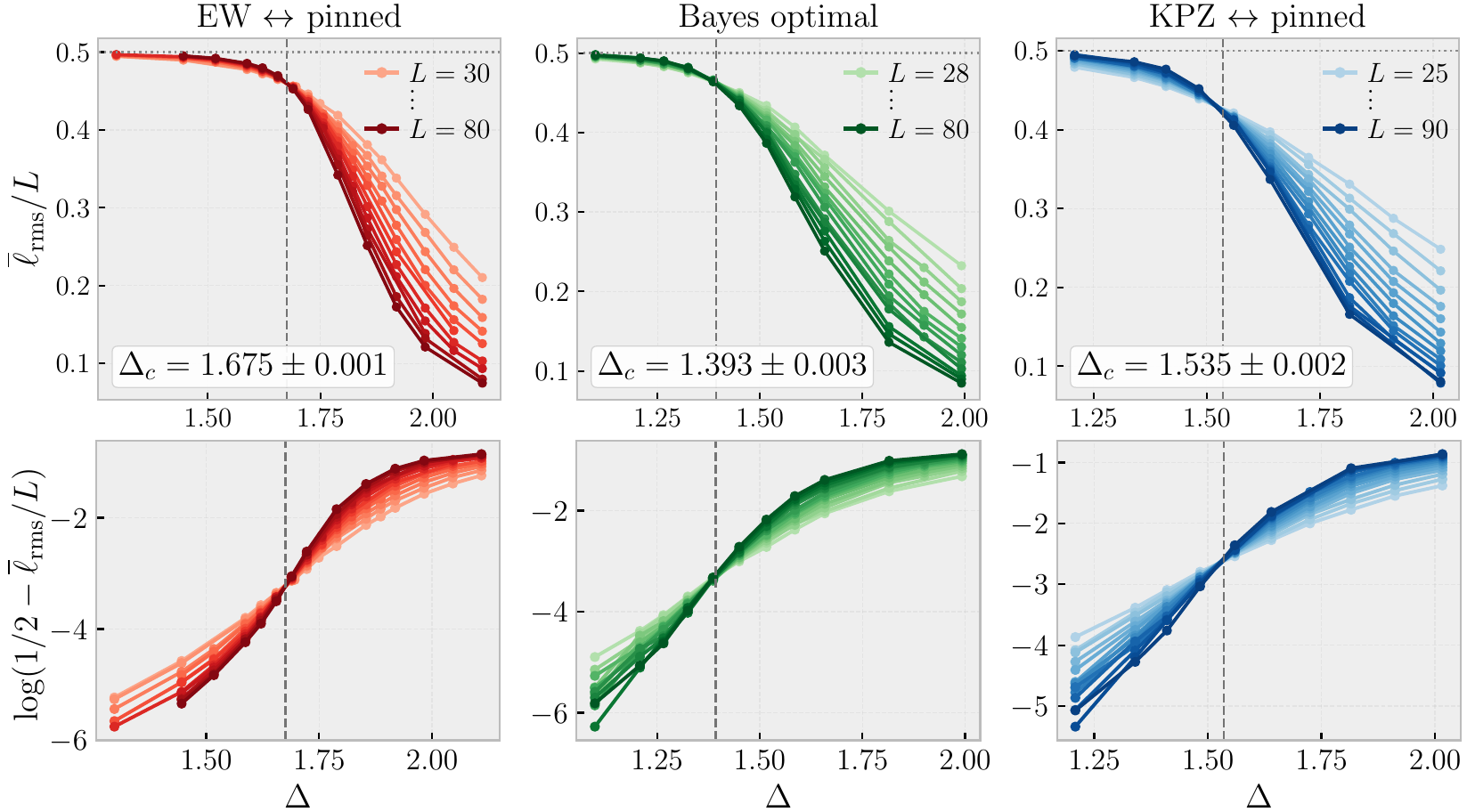}
    \caption{
        RMS distance
        $\overline{\ell}_{\rm rms}/L$ and transformed variable $Y_L(\Delta) = \log(1/2-\overline{\ell}_{\rm rms}/L)$ for the EW--to--pinned,
        Bayes-optimal, and KPZ--to--pinned scans in $d=3$.
        The dotted horizontal line marks the delocalized limiting value
        $1/2$, while the dashed vertical lines show the estimates of
        $\Delta_c$ obtained from the logarithmic crossing analysis.
        The compression of the curves near $1/2$ motivates the transformed
        observable shown in the lower panels.
    }
    \label{fig:3d-raw-width}
\end{figure}

\begin{figure}[t!]
    \centering
    \includegraphics[width=0.72\linewidth]
        {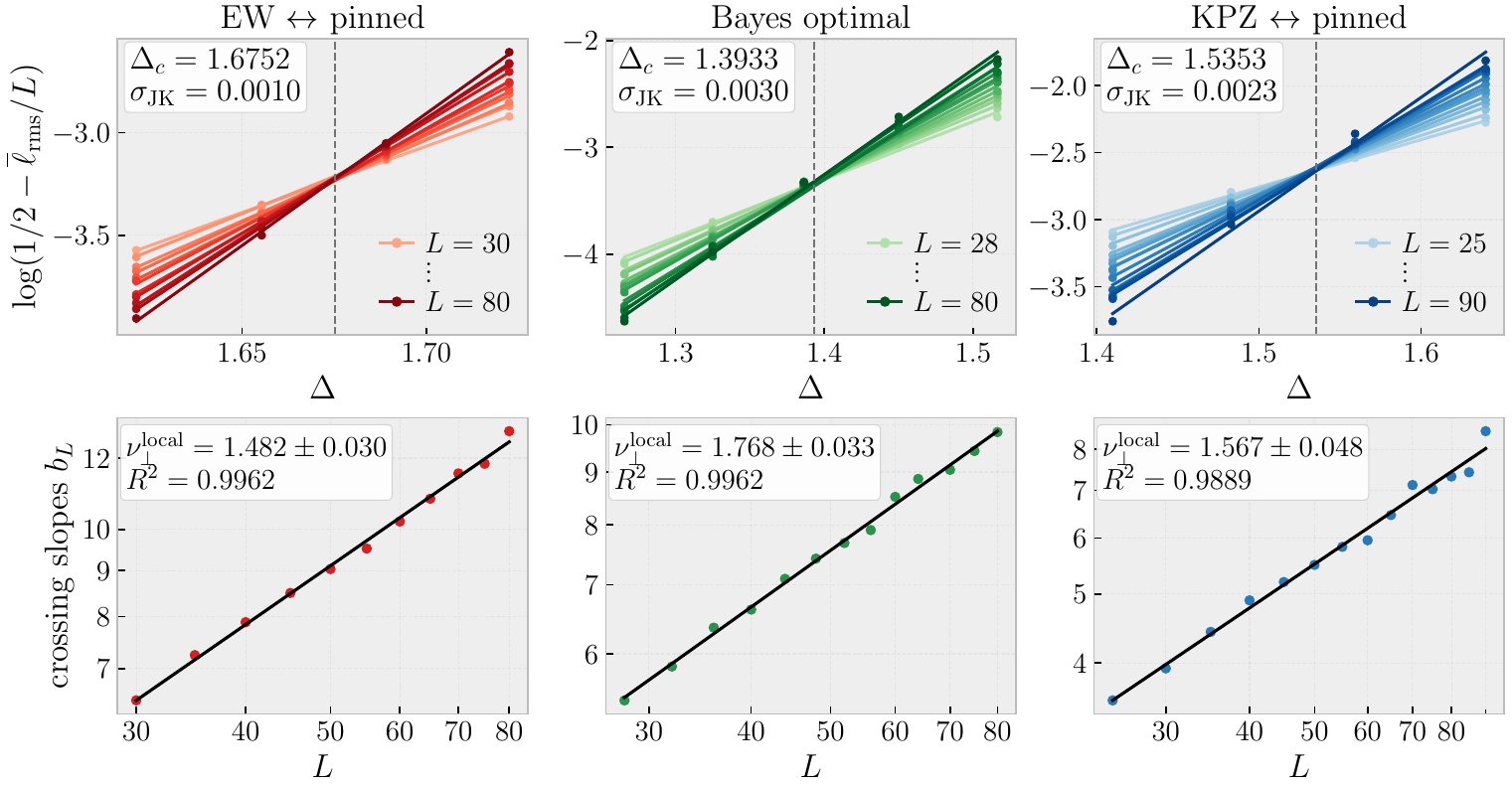}
    \caption{
        Extraction of the $d=3$ depinning thresholds and transverse
        correlation-length exponents.
        Top row: the transformed observable
        $Y_L=\log[1/2-\overline{\ell}_{\rm rms}/L]$ in the local fitting
        window. Solid lines are linear fits $Y_L(\Delta)\simeq a_L+b_L\Delta$, and dashed vertical lines mark the
        common-crossing estimates from $\Delta_c=\mathop{\rm arg\,min}_{\Delta} {\rm Var}_L\!\left(a_L+b_L\Delta\right)$.
        Bottom row: the crossing slopes $b_L$. Black lines
        are fits to $b_L\propto L^{1/\nu_\perp}$, giving the independent
        local estimates $\nu_\perp^{\rm local}$ shown in each panel.
        The displayed $\sigma_{\rm JK}$ is the leave-one-size-out uncertainty
        in $\Delta_c$.
    }
    \label{fig:3d-crossing-slopes}
\end{figure}

To begin with, we analyze the data under the ``conventional'' assumption that the transitions are all associated with crossings at a finite point $(\Delta_c, Y_c)$. Later we will reexamine the data in the light of the theoretical suggestion in Sec.~\ref{sec:infinite-order-transitions} that two of these transitions are instead infinite order transitions.

To locate each transition, we fit $Y_L(\Delta)\simeq a_L+b_L\Delta$ over a restricted set of points around the crossing. We then estimate the crossing as the value of $\Delta$ that minimizes the variance across the fitted lines, $\Delta_c=\mathop{\rm arg\,min}_{\Delta} {\rm Var}_L\!\left(a_L+b_L\Delta\right)$. (Uncertainties in $\Delta_c$ are estimated by repeating this fitting procedure with one $L$ removed.) The upper panels of Fig.~\ref{fig:3d-crossing-slopes} show the local fits and the estimated crossing (and uncertainties).

The local slopes at the crossing provide one estimate of the transverse correlation-length exponent through $|b_L|\sim L^{1/\nu_\perp}$. The uncertainty is then estimated by the regression uncertainty from the straight-line fit of $\log|b_L|$ against $\log L$. The lower panels of Fig.~\ref{fig:3d-crossing-slopes} show an approximately linear relation between $\log|b_L|$ and $\log L$. The data are well described by a finite power law over the available range of sizes, with no evident drift of $1/\nu_\perp$ towards zero (e.g., as one would expect for an infinite-order phase transition with $\nu_\perp = \infty$).

We obtain a second estimate of $\nu_\perp$ by seeking the best collapse of the full finite-size scaling form
\begin{equation}
    Y_L(\Delta)
    =
    \mathcal{Y}\!\left[
        (\Delta-\Delta_c)L^{1/\nu_\perp}
    \right].
    \label{eq:Y-collapse}
\end{equation}
For each trial value of $\nu_\perp$, we rescale the horizontal coordinate as $X=(\Delta-\Delta_c)L^{1/\nu_\perp}$ and use piecewise-cubic interpolation to estimate the common scaling function $\mathcal{Y}(X)$. We quantify the quality of the collapse by the mean squared deviation of the rescaled data from this curve within a fixed window around the transition, and choose $\nu_\perp^{\rm coll}$ to minimize this deviation. As for the crossings, an estimate for uncertainty in each $\nu_\perp$ is obtained by repeating the fitting procedure with one $L$ removed from the data set. Figure~\ref{fig:3D_rms_extra} shows the collapses of the RMS length, as well as the estimated correlation length exponents (and uncertainties).

The contact fraction is fitted for the same data using the scaling form
\begin{equation}
    \overline{ \mathcal P}(0)
    =
    L^{-x}
    \mathcal{Q}\big[
        (\Delta-\Delta_c)L^{1/\nu_\perp}
    \big].
    \label{eq:contact-collapse-app}
\end{equation}
The resulting contact-fraction collapses
are shown in Fig.~\ref{fig:3D_rms_extra}.

\begin{figure}[t!]
    \centering
    \includegraphics[width=0.71\linewidth]
        {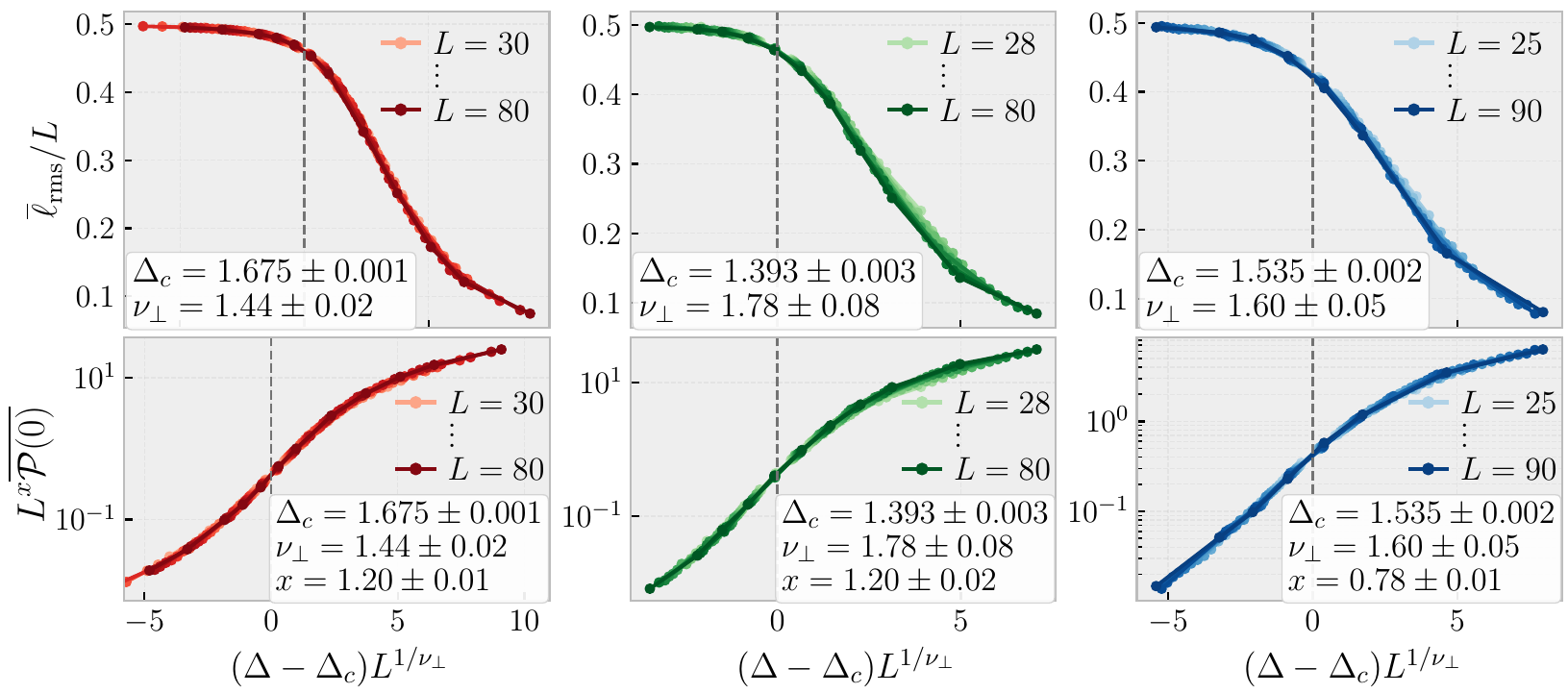}
    \caption{
        Finite-size scaling collapses of
        $\overline{\ell}_{\rm rms}/L$ (top) and the contact fraction $\overline{\mathcal{P}}(0)$ (bottom) for the three $d=3$ diagnostic scans. From left to right, the columns show the EW--to--pinned, Bayes-optimal, and KPZ--to--pinned transitions. The scaling variable is $(\Delta-\Delta_c)L^{1/\nu_\perp}$, using the collapse estimates $\Delta_c$ and $\nu_\perp^{\rm coll}$ listed in Table~\ref{tab:3d-fit-summary}.
    }
    \label{fig:3D_rms_extra}
\end{figure}

Table~\ref{tab:3d-fit-summary} summarizes the crossing locations, the two estimates of $\nu_\perp$, and the contact exponent $x$. The two procedures give mutually consistent estimates of $\nu_\perp$ for all three transitions. The quoted uncertainties characterize sensitivity to the fitting protocol and to the range of finite system sizes. 

These fits assume conventional power-law finite-size scaling with finite $\nu_\perp$, and with a finite value of $Y$ at the transition.
We expect that this holds for the KPZ--to--pinned transition.
However, the arguments in Sec.~\ref{sec:infinite-order-transitions} suggest that the other two transitions have a different behavior, with the candidate polymer being asymptotically Brownian at the transition points, and plausibly with exponent $\nu_\perp=\infty$.
In this scenario, we would expect to see the crossing points drift upwards to ${Y = - \infty}$, so that the value at the critical point matched the unpinned (Brownian) value. 
While the available system sizes span only a factor of $\approx 2.7$, it is nevertheless striking that we do not see any such  tendency of the crossings to drift.
If the correlation length exponent $\nu_\perp$ is infinite, then we would expect to see the curves in Fig.~\ref{fig:3d-crossing-slopes}, lower-left and lower-middle panels, bend downwards, so that the asymptotic gradient of the curves was zero.
Again it is striking that no such tendency is apparent in the data.
This tension between the data and the theoretical arguments is unresolved.

\begin{table}[t!]
    \centering
    \begin{tabular}{lcccc}
\hline
Transition & $\Delta_c$ & $\nu_\perp^{\rm local}$ & $\nu_\perp^{\rm coll}$ & $x$ \\
\hline
EW--pinned & $1.675\pm0.001$ & $1.482\pm0.030$ & $1.440\pm0.017$ & $1.203\pm0.010$ \\
Bayes optimal & $1.393\pm0.003$ & $1.768\pm0.033$ & $1.776\pm0.076$ & $1.196\pm0.017$ \\
KPZ--pinned & $1.535\pm0.002$ & $1.567\pm0.048$ & $1.603\pm0.049$ & $0.779\pm0.015$ \\
\hline
\end{tabular}
    \caption{
        Numerical estimates for the three $d=3$ depinning transitions.
        $\nu_\perp^{\rm local}$ is obtained from fitting crossing slopes as shown in Fig.~\ref{fig:3d-crossing-slopes}, while
        $\nu_\perp^{\rm coll}$ is obtained from the full collapse in
        Eq.~\eqref{eq:Y-collapse}. The uncertainties in $\Delta_c$,
        $\nu_\perp^{\rm coll}$, and $x$ are estimated by repeating the fitting procedures with one $L$ removed from the dataset; whereas in $\nu_\perp^{\rm local}$ the uncertainty is the
        regression uncertainty of the log--log slope fit.
    }
    \label{tab:3d-fit-summary}
\end{table}

\subsection{Misspecification scan}
\label{app:misspecification-numerics}

For the misspecification scan discussed in the main text, we fix the true model and vary only the measurement strength assumed by the observer, $\Delta'=\kappa\Delta$. The true parameters are $(D,p_0,p_1)=(0.0167,0.5,0.76)$, with $p'_0=p_0$, while $p'_1$ is varied through $\kappa$. We use approximately $2\times10^5$ samples for each pair $(L,\kappa)$. The Bayes-optimal point is $\kappa=1$. Moving to $\kappa<1$ makes the observer underconfident and eventually drives the posterior into the EW-unpinned phase, whereas $\kappa>1$ makes the observer overconfident and eventually drives it into the KPZ-unpinned phase.

Figure~\ref{fig:log-score} compares the contact fraction with the expected log score of the true position. The latter is shown through its exponential, $\exp\!\big[\,\overline{\log \mathcal{P}(0)}\,\big]$, which is a monotone function of the expected log score. Bayes-optimal inference maximizes the expected log score, rather than the mean contact fraction itself~\cite{Gneiting2007,Good1952}. Correspondingly, the log-score curve has a clear maximum at $\kappa=1$. The contact fraction instead reaches its maximum slightly on the overconfident side in the pinned phase. Thus maximizing the mean posterior weight at the true position is not, in general, equivalent to choosing the statistically optimal model. The corresponding RMS distance across this scan is shown in Fig.~\ref{fig:3D_main}(d) in the main text.

\begin{figure}[t!]
    \centering
    \includegraphics[width=0.6\linewidth]
        {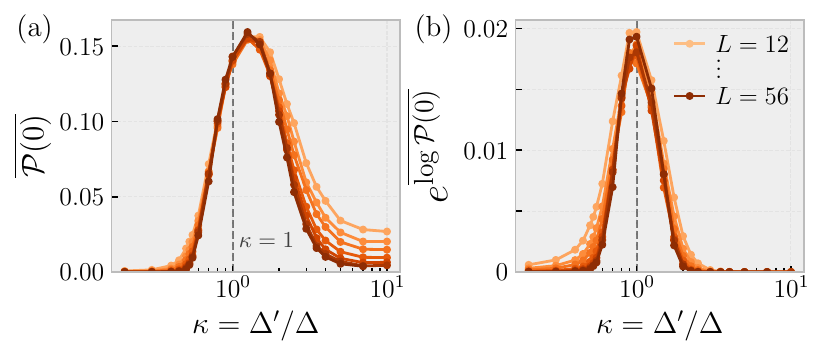}
    \caption{
        Scoring observables along the $d=3$ misspecification scan.
        The true model is fixed at
        $(D,p_0,p_1)=(0.0167,0.5,0.76)$, while the observer assumes
        $\Delta'=\kappa\Delta$ and $p'_0=p_0$.
        The dashed line marks Bayes optimality, $\kappa=1$.
        The expected log score, shown as
        $\exp[\langle\log \mathcal P(0)\rangle]$, is maximized at Bayes optimality,
        whereas the contact fraction $\langle \mathcal P(0)\rangle$ reaches its
        maximum slightly in the overconfident regime. Curves show system
        sizes $L=12,\ldots,56$.
    }
    \label{fig:log-score}
\end{figure}

\section{Bayes-optimal free-energy identity}
\label{app:BO_free_energy}

In this appendix we derive the free-energy derivative identity Eq.~\ref{eq:BO_free_energy_identity} used in Sec.~\ref{sec:3dtopology}. For convenience we  work with the continuum replica description of Eq.~\eqref{eq:replica_polymer_action}, though analogous identities can be obtained on the lattice.
For convenience we will  derive the identities using the replica trick,
but again this is only for convenience, and they can also be obtained without using a replica limit.
Recall that the teacher has replica index $a=0$, while the $n-1$ student trajectories have indices $a=1,\ldots,n-1$, with the replica limit $n\to1$.

For two trajectories define their integrated contact time
\begin{equation}
    N_{ab}
    \equiv
    \int_0^t d\tau\,
    \delta^{(d)}
    \left(
        x_a(\tau)-x_b(\tau)
    \right).
    \label{eq:app_Nab}
\end{equation}
The student--student interaction in the replicated action is therefore $-g_{ss}\sum_{1\leq a<b\leq n-1}N_{ab}$, while the teacher--student interaction is $-g_{st}\sum_{a=1}^{n-1}N_{0a}$.

The Gaussian bulk-disorder average also generates the trivial replica-diagonal contribution $g_{ss}t/2$ to $\overline{\log \mathcal Z_t}$, or equivalently $-g_{ss}t/2$ to the quenched free energy. We remove this analytic contribution by defining
\begin{equation}
    \hat F(g_{ss},g_{st})
    \equiv
    -\overline{\log\mathcal Z_t}
    +
    \frac{g_{ss}t}{2}.
    \label{eq:app_Fe_definition}
\end{equation}
This shift changes only the normalization of the polymer partition function and has no effect on the posterior measure or on the singular part of the free energy.

Let $\mathcal R_n$ denote the normalized disorder-averaged replicated partition function obtained from $\overline{\mathcal Z_t^{\,n-1}}$ after removing the trivial replica-diagonal contribution, so that $\mathcal R_n$ is given explicitly by the replica-polymer path sum in Eq.~\eqref{eq:replica-polymer-path-sum}. Then
\begin{equation}
    \hat F
    =
    -\lim_{n\to1}
    \frac{\mathcal R_n-1}{n-1}.
    \label{eq:app_replica_Fe}
\end{equation}
Differentiating $\mathcal R_n$ with respect to $g_{ss}$ gives
\begin{align}
    \frac{\partial\mathcal R_n}{\partial g_{ss}}
    =
    \mathcal R_n
    \left\langle
        \sum_{1\leq a<b\leq n-1}N_{ab}
    \right\rangle_n =
    \mathcal R_n
    \frac{(n-1)(n-2)}{2}
    \left\langle N_{12}\right\rangle_n ,
    \label{eq:app_dQ_dgss}
\end{align}
where we used the permutation symmetry among the students in the second equality. Taking the replica limit gives
\begin{equation}
    2\frac{\partial \hat F}{\partial g_{ss}}
    =
    \overline{\left\langle N_{12}\right\rangle}.
    \label{eq:app_dFe_dgss}
\end{equation}
Here the right-hand side is the average contact time between two independent posterior trajectories sampled at fixed tracking task, followed by the average over tracking tasks.

Similarly,
\begin{equation}
    \frac{\partial\mathcal R_n}{\partial g_{st}}
    =
    \mathcal R_n (n-1)
    \left\langle N_{01}\right\rangle_n ,
\end{equation}
and hence
\begin{equation}
    -\frac{\partial \hat F}{\partial g_{st}}
    =
    \overline{\left\langle N_{01}\right\rangle},
    \label{eq:app_dFe_dgst}
\end{equation}
where $N_{01}$ is the integrated contact between the teacher and a posterior trajectory.

At Bayes optimality, $g_{ss}=g_{st}$, the $S_{n-1}$ permutation symmetry among the students is enhanced to the full $S_n$ symmetry that exchanges the teacher with the students. Equivalently, the teacher trajectory and an independent trajectory drawn from the posterior are statistically exchangeable. The Nishimori identity therefore gives
\begin{equation}
    \overline{\left\langle N_{01}\right\rangle}
    =
    \overline{\left\langle N_{12}\right\rangle}.
    \label{eq:app_Nishimori_contacts}
\end{equation}
Combining Eqs.~\eqref{eq:app_dFe_dgss},
\eqref{eq:app_dFe_dgst}, and
\eqref{eq:app_Nishimori_contacts}, we obtain
\begin{equation}
    -\frac{\partial \hat F}{\partial g_{st}}
    =
    2\frac{\partial \hat F}{\partial g_{ss}},
    \qquad
    g_{st}=g_{ss}.
    \label{eq:app_BO_derivative_identity}
\end{equation}

\section{Repairing non-optimal inference}

We now give more detail on repairing non-optimal inference (as mentioned in the main text). In the present setting, the inferrer can in principle detect whether or not they are in the pinned (successful) phase  by examining the behavior of the  posterior distribution $\mathcal P(x)$. For example, if $\mathcal P(x)$ is typically localized (detected e.g. by $\ell_{\rm ipr}$ remaining $\mathcal{O}(1)$ at large times),  \textit{and} the localization point moves  diffusively over time in the lab frame  (rather than moving superdiffusively, as would happen in the KPZ-unpinned phase), then this testifies to tracking success. 

Therefore, if the true parameters ${(D, p_0, p_1)}$ are such that Bayes-optimal tracking would be successful, the inferrer can scan $(D', \Delta')$ until the signal tells them that they are in the pinned phase. On the other hand if   ${(D, p_0, p_1)}$ are such that even Bayes-optimal tracking would not succeed, then scanning over suboptimal values of the assumed parameters cannot help. 

Note that it is not necessary for the inferrer to learn the precise values of the true $D$ and $\Delta$ in order to access the pinned phase. The phase diagrams for nonoptimal inference show that successful tracking is stable to nonzero values of ${(D'-D, \Delta'-\Delta)}$ if they are small enough. Roughly speaking, the phase diagram for \textit{non}optimal inference, and the ``size'' of the pinned phase within this phase diagram, determines how easy the scanning task is.

We have assumed here that the inferrer knows the basic structure of the model  and lacks knowledge only of $D'$ and $\Delta'$. (Recall from Eq.~\ref{eq:Zupdate} that $p_0'$ and $p_1'$ appear in the Bayesian update only via the combination in $\Delta'$.) As the amount of prior knowledge about the model is decreased, the scanning task becomes harder in practice, but similar logic can be applied.

\end{document}